\documentclass[12pt,preprint]{aastex}
\usepackage{graphicx}
\begin{document}

%%%%%%%%%%%%%%%%%%%%%%%%%%%%%%%%
%
% Title & Authors
%
%%%%%%%%%%%%%%%%%%%%%%%%%%%%%%%%

\title{The Effect of Variability on the Estimatation of Quasar Black Hole Masses}

\author{
Brian C. Wilhite\altaffilmark{1,2},
Robert J. Brunner\altaffilmark{1,2}, 
Donald P. Schneider\altaffilmark{3}, and
Daniel E. Vanden Berk\altaffilmark{3}
%Jonathan V Brinkmann\altaffilmark{3}
}

\altaffiltext{1}{The University of Illinois, Department of Astronomy,
  1002 W. Green St., Urbana, IL 61801}
\altaffiltext{2}{National Center for Supercomputing Applications, MC-257, 1205 W. Clark Street, Urbana, IL 61801}
\altaffiltext{3}{The Pennsylvania State University, Department of Astronomy and Astrophysics, 525 Davey Lab, University Park, PA 16802}

%\altaffiltext{3}{Apache Point Observatory, P.O. Box 59, Sunspot, NM 88349}

%%%%%%%%%%%%%%%%%%%%%%%%%%%%%%%%
%
% Abstract
%
%%%%%%%%%%%%%%%%%%%%%%%%%%%%%%%%

\begin{abstract}

We investigate the time-dependent variations of ultraviolet (UV) black hole mass estimates of quasars in the Sloan Digital Sky Survey (SDSS).  From SDSS spectra of 615 high-redshift ($1.69 < z < 4.75$) quasars with spectra from two epochs, we estimate black hole masses, using a single-epoch technique which employs an additional, automated night-sky-line removal, and relies on UV continuum luminosity and C~\textsc{iv} $\lambda{1549}$ emission line dispersion.  Mass estimates show variations between epochs at about the $30\%$ level for the sample as a whole.  We determine that, for our full sample, measurement error in the line dispersion likely plays a larger role than the inherent variability, in terms of contributing to variations in mass estimates between epochs.  However, we use the variations in quasars with $r$-band spectral signal-to-noise ratio greater than 15 to estimate that the contribution to these variations from inherent variability is roughly $20\%$.  We conclude that these differences in black hole mass estimates between epochs indicate variability is not a large contributer to the current factor of two scatter between mass estimates derived from low- and high-ionization emission lines.

\end{abstract}


%%%%%%%%%%%%%%%%%%%%%%%%%%%%%%%%
%
% Introduction
%
%%%%%%%%%%%%%%%%%%%%%%%%%%%%%%%%

\section{Introduction\label{intro}}

In active galactic nuclei (AGN) and quasars, it is now generally accepted that for low-ionization broad emission lines, such as H$\beta$, the line width is mostly controlled by gravity, and thus closely related to the mass of the quasar central black hole \citep{osterbrock82,peterson99,peterson00,onken02,kollatschny03}.  However, there is some debate as to the physical processes responsible for producing the observed profile of higher ionization lines, such as C~\textsc{iv} $\lambda$1549.  The C~\textsc{iv} line has been observed to be both blueshifted \citep[see, e.g.][]{wilkes84, richards02b} and asymmetric \citep{wilkes84,vandenberk01} hinting that physical processes other than gravity may be at least partially responsible for the C~\textsc{iv} profile.  Recently, \citet{baskin05} demonstrated that, due to these differences in line profile, black hole mass estimates involving C~\textsc{iv} line width may be less accurate than previously believed, or even biased, perhaps with systematic over or underestimates of mass by a factor of a few.

If one does assume that the width of a given quasar broad emission line can be related to the velocity of gas in orbit around the central black hole, a black hole mass can be computed via the virial equation:

\begin{equation}
M_{BH} = f\frac{R (\Delta{v})^{2}}{G},
\end{equation}
where $f$ is a scale factor of order unity that depends on the geometry of the broad line region (BLR), $R$ is the distance from the black hole to the specific portion of the BLR which contains the emitting gas in question (a distance which likely differs for each species), and $\Delta{v}$ is the velocity width of the broad emission line itself.  While line width is easily determined from a single-epoch spectrum, determining the BLR size is less straightforward.  Reverberation mapping techniques have proven very successful in estimating BLR sizes, and by extension, determining the masses of the black holes at the centers of a few dozen active galactic nuclei \citep[e.g.,][]{wandel99,peterson04}.   These radii are found by measuring the response time of variations in emission line flux to changes in continuum flux.  Though simple in principle, measuring these response times requires constant spectral monitoring, and is observationally taxing.  However, \citet{wandel99} and \citet{kaspi00}, and later \citet{kaspi05}, demonstrated that a simple scaling relationship exists between BLR size and continuum luminosity ($R \propto L^{\alpha}$).   \citet{kaspi05} determined that the size of the H$\beta$ BLR scales with both optical and UV continua, allowing for the use of continuum luminosity in both of these wavelength ranges as a proxy for BLR size and paving the way for reliable, single-epoch black hole mass estimates. 

Single-epoch estimates have been calibrated using the H$\beta$ \citep{wandel99} and Mg~\textsc{ii} $\lambda$2798 \citep{mclure02} emission lines.  \citet{vestergaard02} developed a method of estimating black hole masses derived from C~\textsc{iv} FWHM and $\lambda{\rm L}_{\lambda}$(1350 \AA) (abbreviated L$_{1350}$),  calibrated by the corresponding reverberation-mapping masses, utilizing the scaling relationship determined by \citet{kaspi00}.  Namely, $R_{{\rm C}\textsc{iv}} \propto L_{1350}^{0.70}$.

More recently, \citet{peterson04} reanalyzed a large amount of reverberation mapping data, removing lower-quality data and re-establishing the scaling relationships used for calibration of single-epoch mass estimates.  Subsequently, \citet{kaspi05} used these revised relationships to update the BLR size-continuum luminosity relationships, and \citet{bentz06} made additional corrections after correcting for luminosity contributions from host galaxies' starlight.  These developments led to a re-calibration of UV black hole masses in \citet{vestergaard06}, who utilized an empirically determined radius-luminosity relationship more consistent with photoionization theory: $R_{{\rm C}\textsc{iv}} \propto L_{1350}^{0.53}$.  Early results of a monitoring campaign to apply reverberation mapping to z $\sim3$ quasars indicate that, over 7 orders of magnitude in luminosity, the C~\textsc{iv} BLR size-UV luminosity relationship has a slope similar to that of the H${\beta}$ BLR size-UV luminosity relationship \citep{kaspi07}, confirming the assupmtions made by \citet{vestergaard06}.

Additionally, \citet{vestergaard06} calibrated an estimate for black hole mass which relies upon the dispersion of the C~\textsc{iv} line ($\sigma_{\rm C\textsc{iv}}$, the second moment about the mean), and the luminosity density at 1450\AA:

\begin{equation}
{\rm log\ M_{BH}(C~\textsc{iv})} = {\rm log} \left [ \left ( \frac{{\rm \sigma_{C\textsc{iv}}}}{{\rm 1000\ km\ s^{-1}}} \right )^{2} \left ( \frac{{\rm \lambda{L_{\lambda}(1450\AA)}}}{{\rm 10^{44}\ erg\ s^{-1}}} \right )^{0.53} \right ] + (6.73 \pm 0.01).
\label{vestergaard}
\end{equation}
Based on comparisons with reverberation mapping masses, \citet{vestergaard06} state that these masses are likely good to within a factor of 3.

One potential problem in estimating black hole masses from single-epoch spectra is the inherent variability of quasars.  This variability is key to reverberation mapping techniques, but will necessarily inject uncertainties into single-epoch mass estimates.  The optical and ultraviolet portions of quasar continua have long been known to vary in luminosity on the order of $10\%-20\%$ on time scales from weeks to years \citep[e.g.,][]{smith63,uomoto76,hook94,giveon99,devries03}.  
%{\bf 
\citet{vandenberk04}, using a sample of $\sim25000$ quasars, confirmed known correlations, and parameterized relationships, between variability and rest-frame time lag, luminosity, rest-frame wavelength and redshift.
%}  
Additionally, \citet[][hereafter Paper I]{wilhite05} completed the first study of the detailed dependence of variability upon wavelength, demonstrating that that variablity increased with decreasing wavelength, but only at wavelengths less than 2500\AA.  Increased variability at shorter wavelengths, such as that seen in Paper I, and earlier \citep[e.g.,][]{cutri85,collier01,vandenberk04}, can impact black hole mass estimates that rely on rest-frame UV luminosity.

In addition, the fluxes and profiles of quasar emission lines are known to vary with time \citep[e.g.,][]{peterson93,wanders96,wandel99,sergeev01}, mostly in response to fluctuations in continuum luminosity.  C~\textsc{iv} has been closely monitored in a relatively small number of low-redshift, low-luminosity objects like NGC 5548 \citep{clavel91,korista95} and NGC 4151 \citep{crenshaw96}, as well as in a few high-redshift quasars \citep{kaspi07}.  Recently, \citet[][hereafter Paper II]{wilhite06} studied C~\textsc{iv} variability in an ensemble of $\sim100$ SDSS quasars with multiple-epoch spectroscopy, finding that the width of an individual C~\textsc{iv} line increases with line flux, and varies by as much as 30\%, on rest-frame time scales of weeks to months. Paper I focused on the variability of the quasar continuum, while Paper II centered on variability of the C~\textsc{iv} line.  Given the interest in black hole mass estimates, we feel there is a definite need to re-examine UV variability in the context of mass estimators, and to attempt to quantify the effect  (or lack thereof) of variability on determining black hole masses.

We briefly describe the quasar sample and the additional, necessary spectrophotometric 
calibrations in \S\,\ref{dataset}. We describe the process used to estimate black hole masses, including the continuum- and line-fitting techniques used, in  \S\,\ref{bhmasses}.  The epoch-to-epoch black hole mass estimate differences are examined in \S\,\ref{mbhvar}.  The results are discussed in \S\,\ref{discussion}, and we conclude in \S\,\ref{conclusions}.

For consistency with Papers I and II, we assume a flat, cosmological-constant-dominated cosmology with parameter values $\Omega_\Lambda = 0.7, \Omega_{M} = 0.3,$ and $H_{0}=70$km s$^{-1}$ Mpc$^{-1}$ to calculate luminosity distances.  Though these parameter values differ slightly from recent measurements \citep[e.g.,][]{spergel06}, this should have little effect on results, as we are chiefly concerned with the effect of variability between epochs.

%%%%%%%%%%%%%%%%%%%%%%%%%%%%%%%%
%
% Data
%
%%%%%%%%%%%%%%%%%%%%%%%%%%%%%%%%

\section{The Quasar Dataset\label{dataset}}

\subsection{The Sloan Digital Sky Survey\label{SDSS}}

The Sloan Digital Sky Survey \citep{york00}, using a dedicated 2.5-meter telescope \citep{gunn06} at the Apache Point Observatory in the Sacramento Mountains of New Mexico, has, through Summer 2005, acquired imaging and spectroscopic data for $\sim$8000 deg$^{2}$, mostly centered on the Northern Galactic Cap.  A 54-chip drift-scan camera \citep{gunn98} acquires imaging data which are reduced and calibrated by using the astrometric \citep{pier03} and photometric  \citep{lupton01} software pipelines.  The photometric system is normalized such that SDSS $u,g,r,i$ and $z$ magnitudes \citep{fukugita96} are on the AB system \citep{smith02}.  A 0.5-meter telescope monitors site photometric quality and extinction \citep{hogg01,ivezic04,tucker06}.   

After image processing, selected objects are targeted for spectroscopy \citep{strauss02,eisenstein01,richards02a, stoughton02} and grouped in 3-degree diameter tiles \citep{blanton03}.  For each tile, an aluminum plate is drilled with 640 holes reserved for roughly 500 galaxies, 50 quasars and 50 stars (40 calibration spectra---32 sky fibers and 8 reddening standards---are also taken with each plate).  Plates are placed in the imaging plane of the telescope and the holes plugged with optical fibers running from the telescope to twin spectrographs.

SDSS spectra are obtained in three or four consecutive 15-minute observations and cover the observer-frame optical and near infrared, from 3900\AA--9100\AA.  The {\tt Spectro2d} pipeline flat-fields and flux calibrates spectra, and {\tt Spectro1d} identifies spectral features and classifies objects by spectral type \citep{stoughton02}.  Extragalactic objects with broad emissions lines (FWHM $\gtrsim 1000$km s$^{-1}$) are defined to be quasars.

As we are interested in variations in spectroscopic mass-estimation techniques, we focus here on those quasars that have multiple spectroscopic observations.  Through June 2004, objects on 181 different plates had been observed at least twice, with time lags between observations ranging from days to years.  Fifty-three of these plates (containing roughly 2200 quasars) have observations more than 50 days apart, indicating that spectra from these observations have not been co-added and are, therefore, appropriate for variability studies (see Paper I for a lengthier discussion of these data).  52 of these 53 plate pairs are contained in the Fourth Data Release \citep[DR4;][]{adelman06}.

\subsection{Refinement of Spectroscopic Calibration \label{calib}}

It was demonstrated in both \citet{vandenberk04} and Paper I that additional spectrophotometric calibration beyond the standard SDSS processing is required for variability studies.  Paper I contains a complete discussion of those calibration methods; we briefly summarize the salient points here.  The {\tt Spectro1d} pipeline determines three signal-to-noise (S/N) ratios for each spectrum by calculating the median S/N ratio per pixel in the sections of the spectrum corresponding to the SDSS $g, r$ and $i$ filter transmission curves.  Hereafter, we use the phrase ``high-S/N epoch" to refer to the observation with the higher median $r$-band signal-to-noise ratio and  ``low-S/N epoch"  to refer to the observation with the lower median $r$-band signal-to-noise ratio.  Although most objects follow the plate-wide trend, this does not address the relative S/N values for any given individual object, nor does it correspond to an object's relative line or continuum flux at a given epoch.  The stars on a plate are used to resolve calibration differences between the high- and low-S/N epochs, under the assumption that the majority of stars are non-variable (obviously variable stars are removed from this re-calibration).  For each pair of observations, we create a re-calibration spectrum, equal to the ratio of the median stellar high-S/N epoch flux to the median stellar low-S/N flux, as a function of wavelength.  This re-calibration spectrum is fitted with a 5th-order polynomial to preserve real wavelength dependences, but remove pixel-to-pixel noise (see Figure 5 of Paper I), leaving a smooth, relatively featureless curve as a function of wavelength. All low-S/N epoch spectra are rescaled by this ``correction" spectrum.

\begin{figure}
\includegraphics[scale=1.0]{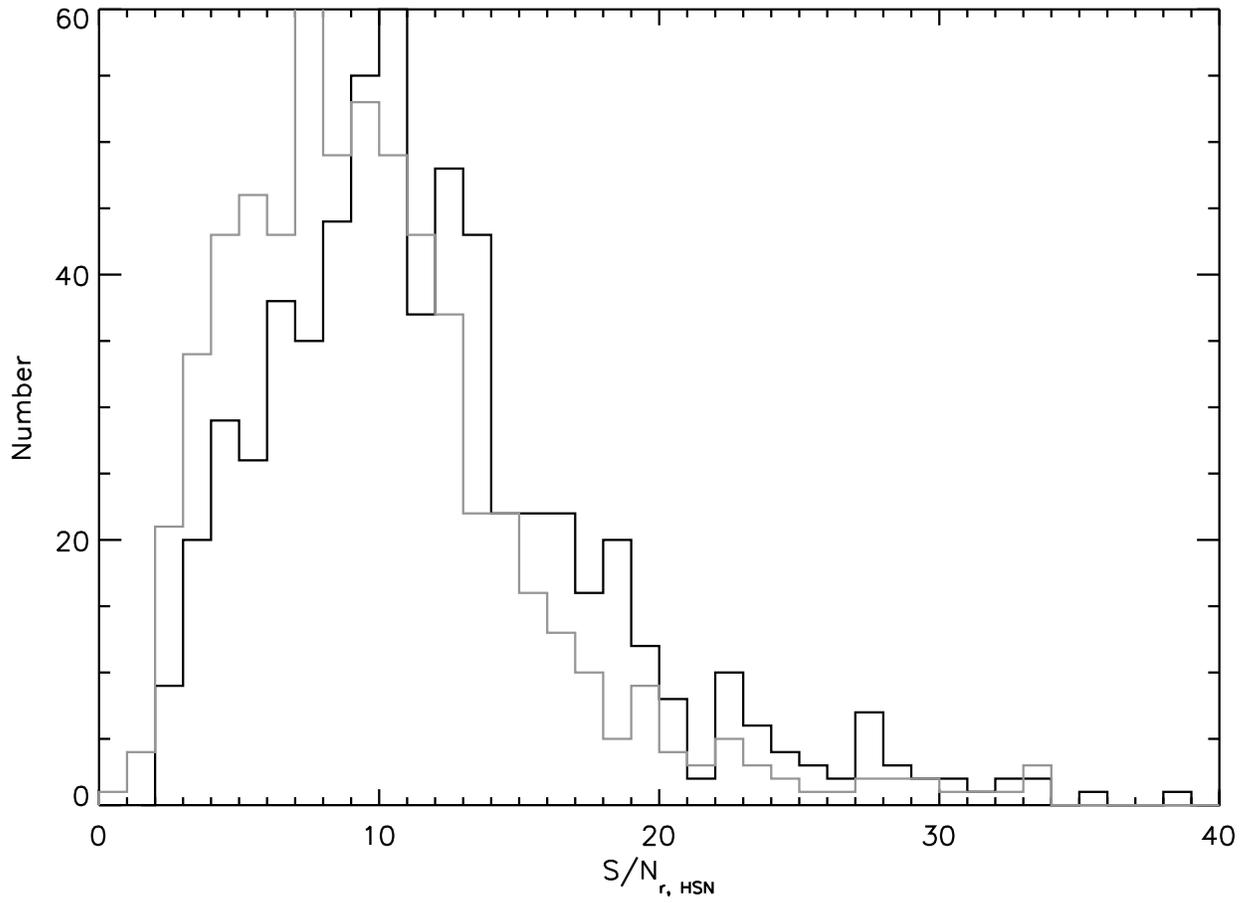}
\caption{$r$-band signal-to-noise ratio at the high-S/N (dark histogram) and low-S/N (gray histogram) epochs.
\label{Fig2.1}}
\end{figure}

\begin{figure}
\includegraphics[scale=1.0]{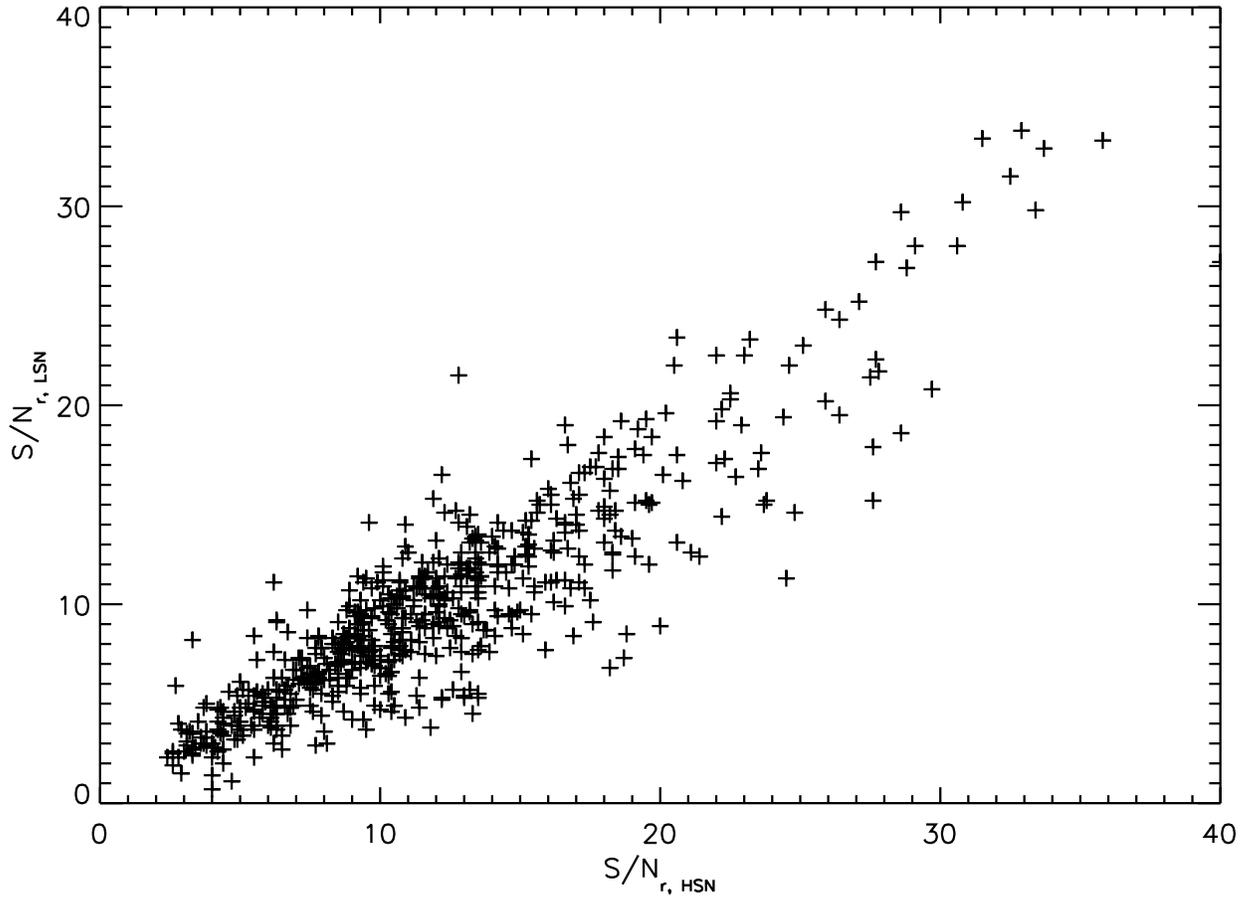}
\caption{$r$-band signal-to-noise ratio at the high-S/N epoch versus the same quantity at the low-S/N epoch.
\label{Fig2.2}}
\end{figure}

\begin{figure}
\includegraphics[scale=1.0]{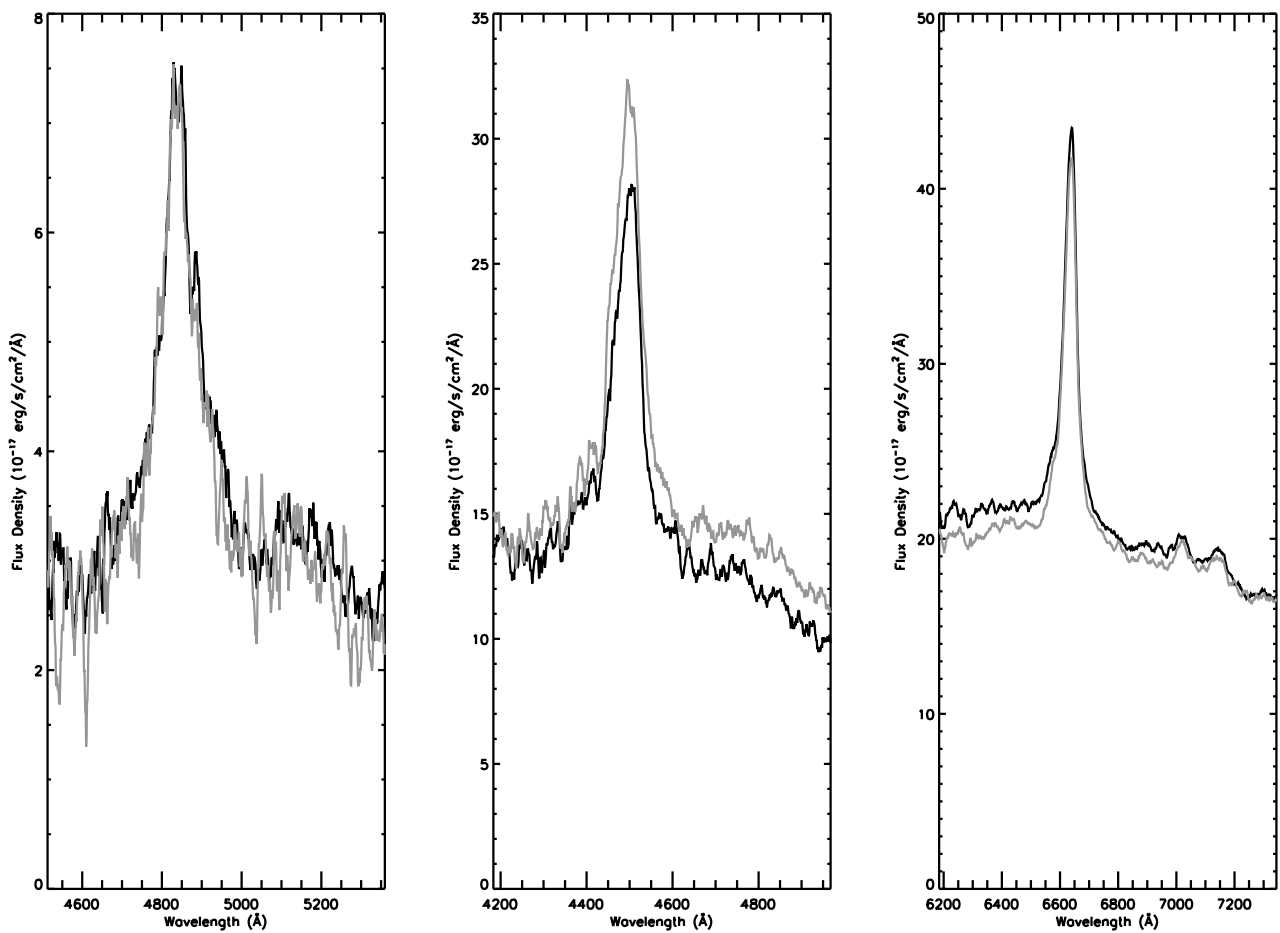}
\caption{Example spectra from the quasars studied in this paper, with objects increasing in spectra signal-to-noise ratio from left to right.  Shown in the observed frame are the regions of the spectra used in the estimation of black hole mass.  Dark curves represent the spectra from the high-S/N epoch, while grey curves are those spectra for the low-S/N epoch. (Left) SDSS J150104.94$-$010727.9; ${\rm S/N}_{r, {\rm HSN}}$=4.9 (Center) SDSS J101416.97+484816.1; ${\rm S/N}_{r, {\rm HSN}}$=12.1 (Right)  SDSS J030449.86$-$000813.4; ${\rm S/N}_{r, {\rm HSN}}$=30.8.
\label{Fig2.3}}
\end{figure}

\begin{figure}
\includegraphics[scale=1.0]{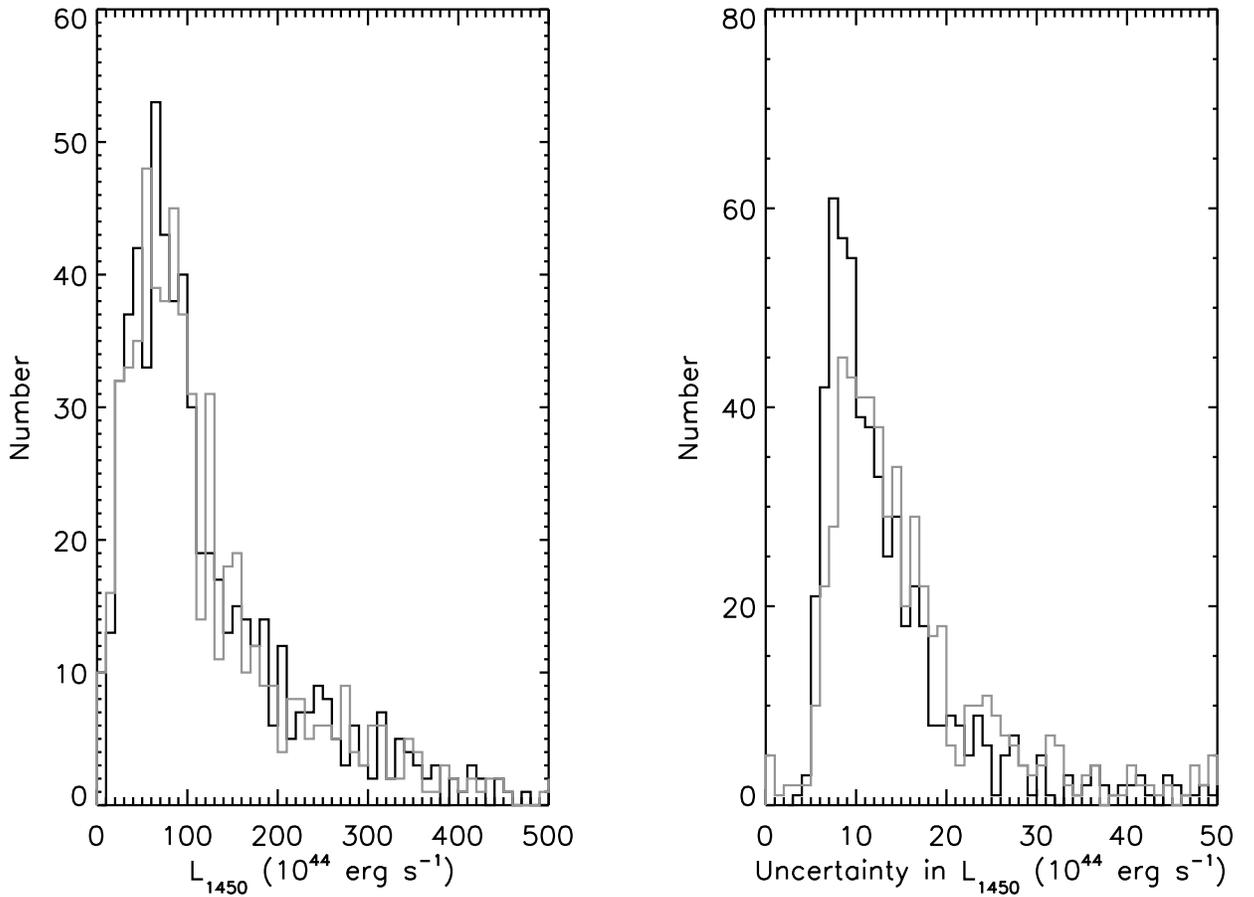}
\caption{(Left) 1450\AA\ luminosity (L$_{1450}$) at the high-S/N (dark histogram) and low-S/N (gray histogram) epochs. (Right) Uncertainty in the 1450\AA\ luminosity at the high-S/N (dark histogram) and low-S/N (gray histogram) epochs.
\label{Fig3.1}}
\end{figure}

In Papers I and II, we studied only those objects that had been shown to vary significantly between epochs.  Here we measure C~\textsc{iv} line width, and estimate the central black hole mass, for all objects in which the entire C~\textsc{iv} line and the 1450\AA\ luminosity are observed.  (As discussed in \S\ \ref{fwhm}, this corresponds to objects with $1.69 < z < 4.75$.)  Out of the main sample of 2210 quasars, 702 are at a redshift where C~\textsc{iv} measurements can be made in the SDSS spectra.  Of these, 87 (13$\%$) are noted in \citet{lundgren06} for showing evidence of broad absorption near the C~\textsc{iv} emission line.  Because of the difficulties broad absorption lines (BALs) can create in estimating the continuum flux and fitting the C~\textsc{iv} emission line these BAL quasars are removed, leaving 615 objects to comprise the main sample studied below.  Table {\ref{tab1} gives a summary of the observations used in this paper, including the names and redshifts of the quasars observed, as well as the Modified Julian Dates (MJDs) and signal-to-noise ratios of the individual observations.

The distributions of the $r$-band spectral signal-to-noise ratio at both epochs are shown as histograms in Figure \ref{Fig2.1}.  The mean ${\rm S/N}_{r}$ at the high-S/N epoch (${\rm S/N}_{r, {\rm HSN}}$) is 12.0, while the mean ${\rm S/N}_{r, {\rm LSN}}=9.9$.  

Figure \ref{Fig2.2} shows ${\rm S/N}_{r, {\rm HSN}}$ versus ${\rm S/N}_{r, {\rm LSN}}$.  As mentioned above, ``high-S/N" or ``low-S/N" epoch is a plate-wide designation; thus a few individual objects actually have greater ${\rm S/N}_{r, {\rm LSN}}$ than ${\rm S/N}_{r, {\rm HSN}}$.  For the vast majority of objects, however ${\rm S/N}_{r, {\rm HSN}} > {\rm S/N}_{r, {\rm LSN}}$.  
In addition, most objects have ${\rm S/N}_{r}$ at the two epochs such that they lie near the line ${\rm S/N}_{r, {\rm HSN}} = {\rm S/N}_{r, {\rm LSN}}$.  Therefore, when examining the effects of spectral signal-to-noise ratio upon variations in black hole mass estimates between epochs, we will rely upon ${\rm S/N}_{r, {\rm HSN}}$.

\subsection{Sample Spectra\label{spectra}}

Figure \ref{Fig2.3} shows observed-frame spectra at both epochs for three quasars from the sample to demonstrate how spectral $r$-band signal-to-noise ratio relates to overall spectral quality.  These three quasars, SDSS J150104.94$-$010727.9 (Quasar 149 in Table 1; ${\rm S/N}_{r, {\rm HSN}}$=4.9), SDSS  J101416.97+484816.1 (Quasar 551; ${\rm S/N}_{r, {\rm HSN}}$=12.1) and SDSS J 030449.86$-$000813.4(Quasar 259; ${\rm S/N}_{r, {\rm HSN}}$=30.8) were chosen to represent a range of ${\rm S/N}_{r, {\rm HSN}}$ values.  Only the region of the spectrum used in the estimation of black hole mass (corresponding to the rest-frame interval [1440\AA,1710\AA]; see \S\ref{bhmasses}) is shown.

%%%%%%%%%%%%%%%%%%%%%%%%%%%%%%%%
%
% Results
%
%%%%%%%%%%%%%%%%%%%%%%%%%%%%%%%%

\section{Calculating Black Hole Mass Estimates\label{bhmasses}}

This paper uses the \citet{vestergaard06} UV black hole mass estimator, seen in Equation 2, which requires measuring the continuum flux at a wavelength blueward of the C~\textsc{iv} line, as well as the dispersion of the C~\textsc{iv} line itself.  The measurements of these two quantities are described below.

\subsection{Sky Subtraction\label{skysub}}

It was determined in Paper II that occasional errors in the SDSS night-sky removal pipeline could lead to errors in continuum and line fitting.  In a small fraction (less than $5\%$) of objects, night sky lines are significantly under- or over-subtracted.  In Paper II, spectra were visually inspected for signs of poor night sky subtraction.  For this work, with over 600 C~\textsc{iv} emission lines to fit (and for future work with larger samples of SDSS quasars), night sky subtraction has been automated. The night sky lines for which the algorithm searches are OI $\lambda5577$, Na $\lambda5890$, OI $\lambda6300$ and the well-known atmospheric O$_{2}$ Fraunhofer A and B bands (covering the [7594\AA, 7621\AA] and [6867\AA, 6884\AA] intervals, respectively).   If any of these known night sky lines lies in the part of the spectrum corresponding to the rest-frame interval [1440\AA, 1700\AA] (the interval used to meausure continuum luminosity and line dispersion; see \S\S\,3.2--3.3), the algorithm tests to ensure that the pipeline night sky subtraction was done properly.  The average flux in a 10\AA\ region centered on the night sky line position (37\AA\ and 27\AA\ regions are used for the wider A and B bands, respectively) is compared to the average flux of the 25\AA\ range on either side of the 10\AA\ region.  If the night sky region flux is more than 3 standard deviations larger or smaller than the average flux of the surrounding region, then the flux in the night sky region is estimated using a linear interpolation based upon the pixels in the surrounding continuum region.  If the 25\AA\ range overlaps with the C~\textsc{iv} emission line (corresponding to the rest-frame interval [1496\AA, 1596\AA]), the region is truncated to include only known continuum flux.  The flux density uncertainties in the individual pixels are not altered, however.  This may lead to a slight overestimation of the errors in the given quantities, but it not likely to have a large effect.

\subsection{1450\AA\ Continuum Luminosity\label{luminosity}}

After the night sky subtraction errors have been corrected, the 1450\AA\ flux density, f$_{\lambda}$(1450), is calculated by taking the mean of the flux density in the pixels corresponding to the rest-frame interval [1445\AA, 1455\AA].  This is translated to a luminosity density by calculating the luminosity distance analytically from the redshift, and then to luminosity by multiplying by wavelength: L$_{1450}$ = 1450\AA\ $\times\ $L$_{\lambda}$(1450).  Figure \ref{Fig3.1} shows the distribution of 1450\AA\ luminosities, L$_{1450}$ at both epochs.  Values for L$_{1450}$ range from 10 to roughly 500 $\times\ 10^{44}$ erg s$^{-1}$, with a median at the high-S/N epoch of 93.1 $\times\ 10^{44}$ erg s$^{-1}$ and 92.6 $\times\ 10^{44}$ erg s$^{-1}$ at the low-S/N epoch.  

The distribution of estimated uncertainties in  L$_{1450}$, calculated through standard error propagation, with the standard deviation in flux in the [1445\AA, 1455\AA]  interval used as the uncertainty in f$_{\lambda}$(1450), is shown for both epochs in Figure \ref{Fig3.1}.  These uncertainties are roughly an order of magnitude smaller than the luminosities themselves, with a median of 11.6 $\times  10^{44}$ erg s$^{-1}$ at the high-S/N epoch and 13.9 $\times\ 10^{44}$ erg s$^{-1}$ at the low-S/N epoch.  
%Figure \ref{Fig3.3} shows the calculated uncertainty at each epoch as a function of that epoch's signal-to-noise ratio.  As one would expect, objects with a low value of ${\rm S/N}_{r}$ have a greater likelihood of a large uncertainty in L$_{1450}$.

\subsection{C~\textsc{IV} Line Dispersion}\label{fwhm}}

The dispersion of the C~\textsc{iv} line is calculated in the same manner as in Paper II.  We briefly describe that procedure here; for a full descrption, see \S3 of that paper.

To avoid contamination from unidentified emission just redward of the C~\textsc{iv} emission line \citep[see, e.g.,][]{wilkes84, boyle90,laor94,vandenberk01}, as well as known emission from He~\textsc{ii}, O~\textsc{iii]}, Al~\textsc{ii]} and Fe~\textsc{ii}, the red side of the continuum is fit over the rest-frame interval [1685\AA, 1700\AA].  The blue continuum is fit over the interval [1472\AA, 1487\AA].  Then, using all pixels corresponding to either of these wavelength ranges, the continuum is fit with a linear least squares algorithm (POLY\_FIT in IDL).  Once the fit has been performed, the continuum fit is subtracted from the entire region of interest. 

To measure the line profile, we integrate over the interval [1496\AA, 1596\AA].  This allows us to exclude potentially contaminating flux blueward of the line profile.  For reasons of stability, we opt to use the line median, rather than the mean, in measuring the line center, as the mean is too easily affected by noisy pixels in the line wings.  
%{\bf 
The median is simply the midpoint of the line flux, the wavelength which evenly divides the continuum-subtracted flux in the line profile.
%}

\begin{figure}
\includegraphics[scale=1.0]{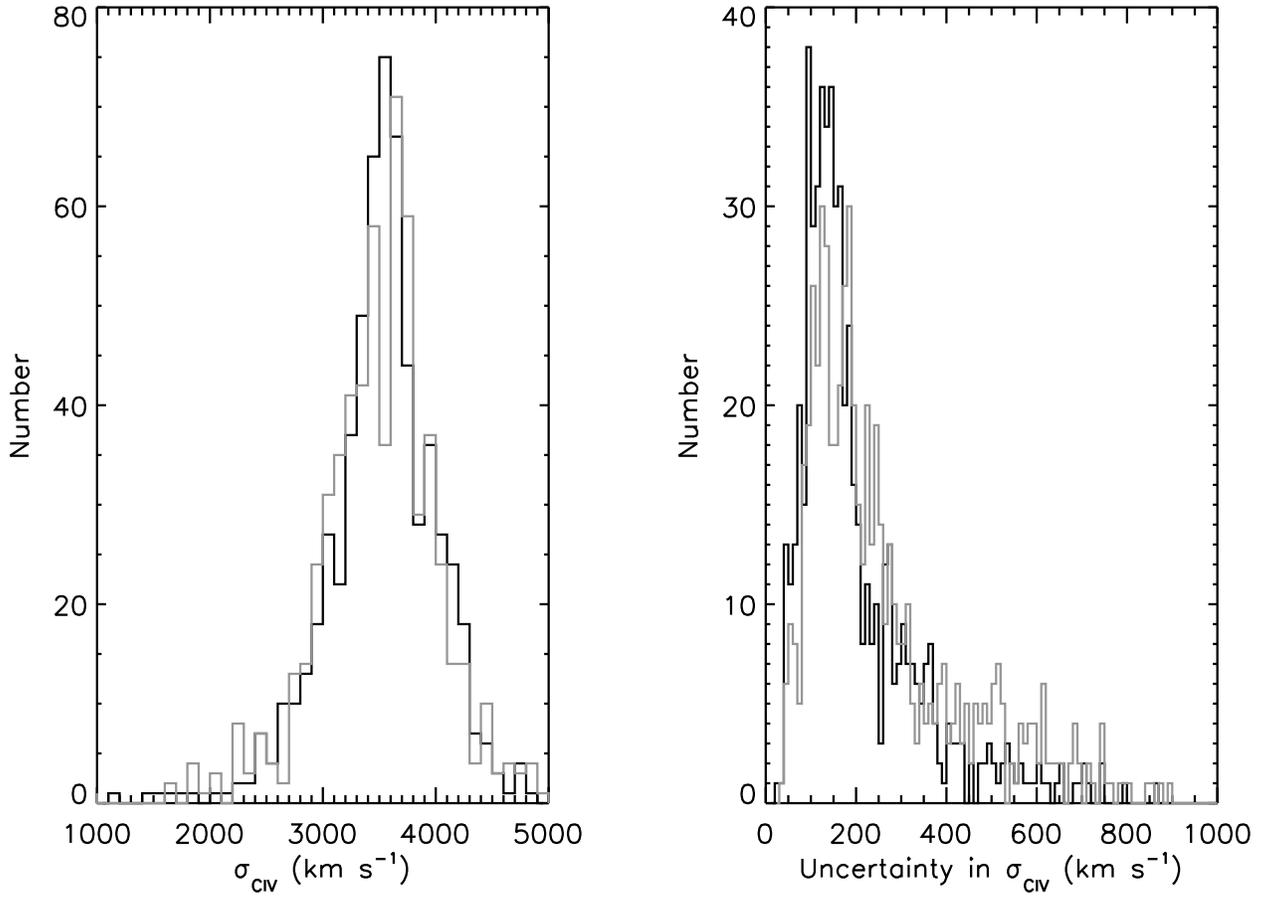}
\caption{(Left) C~\textsc{iv} line dispersion ($\sigma_{C\textsc{iv}}$) at the high-S/N (dark histogram) and low-S/N (gray histogram) epochs.  (Right) Uncertainty in the C~\textsc{iv} line dispersion at the high-S/N (dark histogram) and low-S/N (gray histogram) epochs.
\label{Fig3.4}}
\end{figure}

We then calculate the line profile dispersion, the second moment about the median wavelength:

\begin{equation}
\sigma^{2}=\frac{\int_{\rm {C\textsc{iv}}}(\lambda-\lambda_{median})^{2}F_{\lambda}d\lambda}{\int_{\rm {C\textsc{iv}}}F_{\lambda}d\lambda},
\end{equation}
where C~\textsc{iv} in the integrals simply means we are integrating over the [1496\AA, 1596\AA] interval containing the entire line profile.  Line widths range from 1000 to 5000 km s$^{-1}$, with median values of 3541 km s$^{-1}$ and 3531 km s$^{-1}$ at the high- and low-S/N epochs, respectively.  A histogram of measured line widths for both epochs is shown in Figure \ref{Fig3.4}. 

To determine the uncertainties in these quantities, we use a Monte Carlo method.  Noise is added to each pixel in the region of interest 
%{\bf 
by assigning a random number drawn from a Gaussian distribution with mean equal to the measured flux in that pixel and standard deviation equal to the measured error in that pixel.
%}  
The continuum is fit, and the line median and standard deviation calculated; this is done 1000 times per quasar.  The standard deviation of the distribution of resulting values is assigned to be the error in that quantity.  The uncertainties in the line width (as seen in Figure \ref{Fig3.4}) are, for the most part, less than 500 km s$^{-1}$, with a median uncertainty of 159 km s$^{-1}$ at the high-S/N epoch and 202 km s$^{-1}$ at the low-S/N epoch---like with L$_{1450}$, the uncertainties are roughly an order of magnitude lower than the values themselves.  
%Figure \ref{Fig3.6} shows the uncertainty in line dispersion as a function of ${\rm S/N}_{r}$ at each epoch.  

\subsection{Single-Epoch Mass Estimates\label{mbhfull}}

\begin{figure}
\includegraphics[scale=1.0]{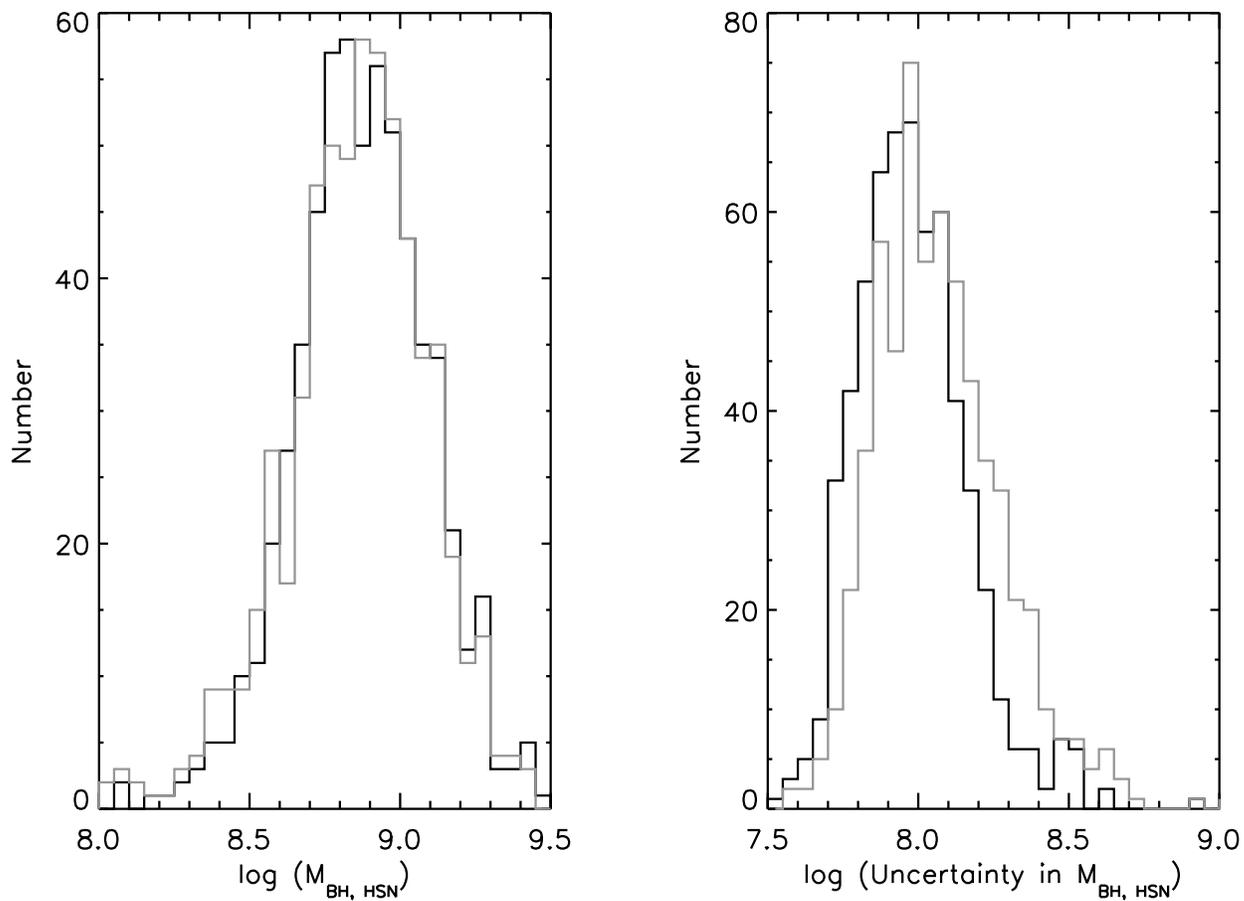}
\caption{(Left) Logarithm of the estimated black hole mass (M$_{BH}$) at high-S/N (dark histogram) and low-S/N (gray histogram) epochs. (Right) Logarithm of the uncertainty ((calculated by propagating measurement errors in L$_{1450}$ and ${\rm \sigma_{C\textsc{iv}}}$) in the estimated black hole mass at the high-S/N (dark histogram) and low-S/N (gray histogram) epochs.
\label{Fig3.7}}
\end{figure}

Once the continuum luminosity (L$_{1450}$) and emission line dispersion (${\rm \sigma_{C\textsc{iv}}}$) have been calculated, it is straightforward to estimate the quasar's black hole mass from Equation 2.  This is done at both the high- and low-S/N epochs for all 615 objects in the main sample.  The distributions of high- and low-S/N-epoch black hole masses are shown in Figure \ref{Fig3.7}.  The majority of objects are estimated to have high-S/N black hole masses in the range from $10^{8.5}$ M$_{\sun}$ to $10^{9.5}$ M$_{\sun}$.  The median high-S/N and low-S/N-epoch masses are $10^{8.88}$ M$_{\sun}$ and $10^{8.87}$ M$_{\sun}$, respectively.  The distributions of uncertainties in M$_{BH}$ (calculated by propagating measurement errors in L$_{1450}$ and ${\rm \sigma_{C\textsc{iv}}}$) at each epoch are shown in Figure \ref{Fig3.7}.  The median uncertainty at the high-S/N epoch is $10^{7.97}$ M$_{\sun}$; at the low-S/N epoch it is $10^{8.04}$ M$_{\sun}$.

\begin{figure}
\includegraphics[scale=1.0]{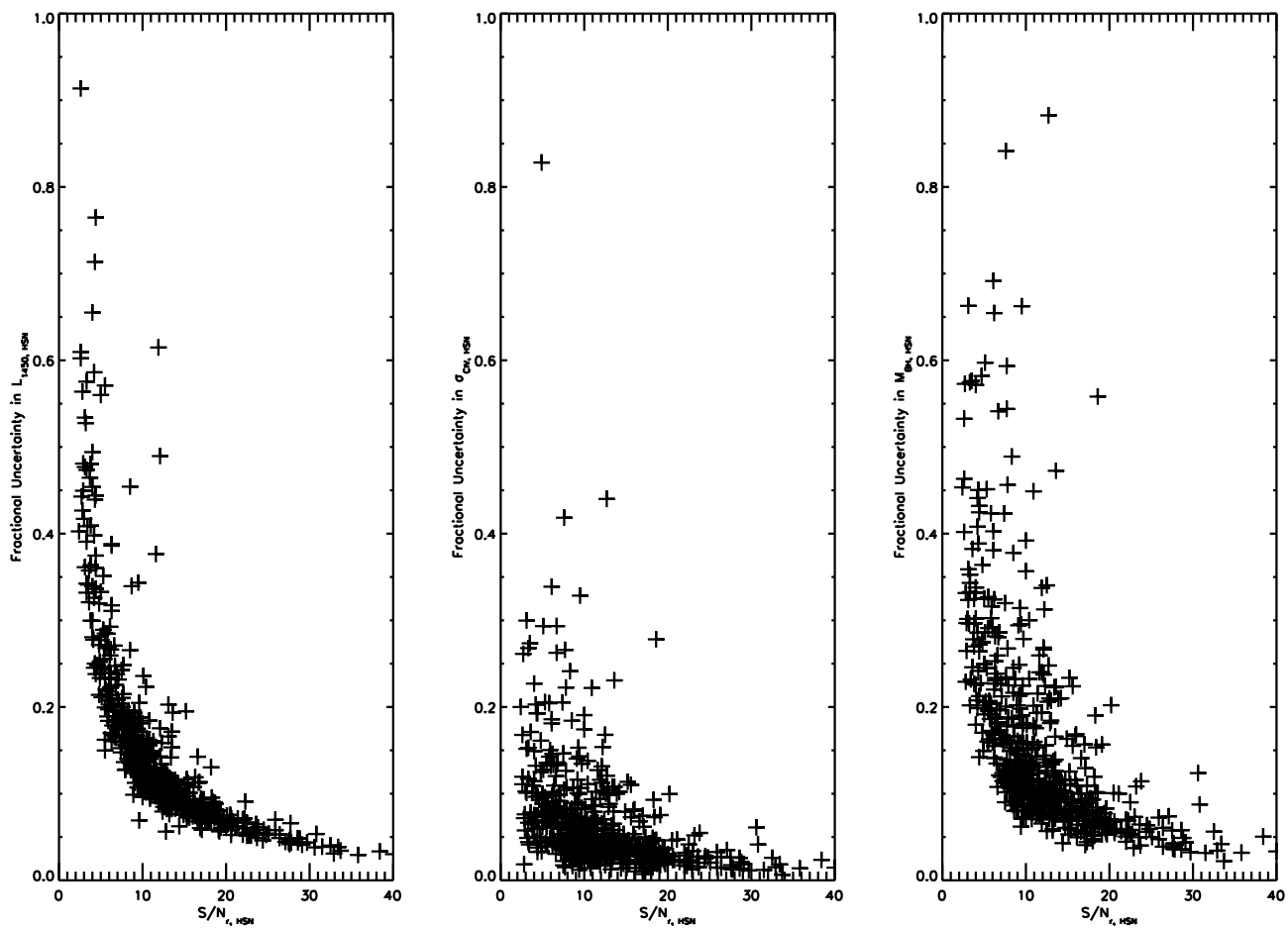}
\caption{(Left) Fractional measurement uncertainty in the 1450\AA\ luminosity ($\sigma_{\rm L_{1450}}/{\rm L_{1450}}$) as a function of $r$-band signal-to-noise ratio (${\rm S/N}_{r}$) at the high-S/N epoch.  (Center) The same, but for uncertainty in the C~\textsc{iv} line dispersion.  (Right) The same, but for uncertainty in the estimated black hole mass.
\label{Fig3.9}}
\end{figure}

Figure \ref{Fig3.9} shows the fractional uncertainty in L$_{1450}$, ${\rm \sigma_{C\textsc{iv}}}$, and M$_{\rm BH}$ as a function of ${\rm S/N}_{r}$ at the high-S/N epoch. (The low-S/N versions of these plots are  very similar and, therefore, not shown.)  The uncertainty in the 1450\AA\ luminosity appears to dominate the M$_{\rm BH}$ measurement error.  It should also be noted that for virtually all quasars with a signal-to-noise ratio greater than 15, our estimate of the measurement uncertainty for M$_{\rm BH}$ is less than $10\%$.

Table \ref{tab2} contains the relevant quantities in the estimation of black hole mass at both epochs, including the masses themselves, and the measured luminosities and line dispersions.

\section{Measuring the Consistency of Estimates of M$_{BH}$\label{mbhvar}}

\subsection{Variations in Luminosity and Line Dispersion\label{lumfwhm_01}}

\begin{figure}
\includegraphics[scale=1.0]{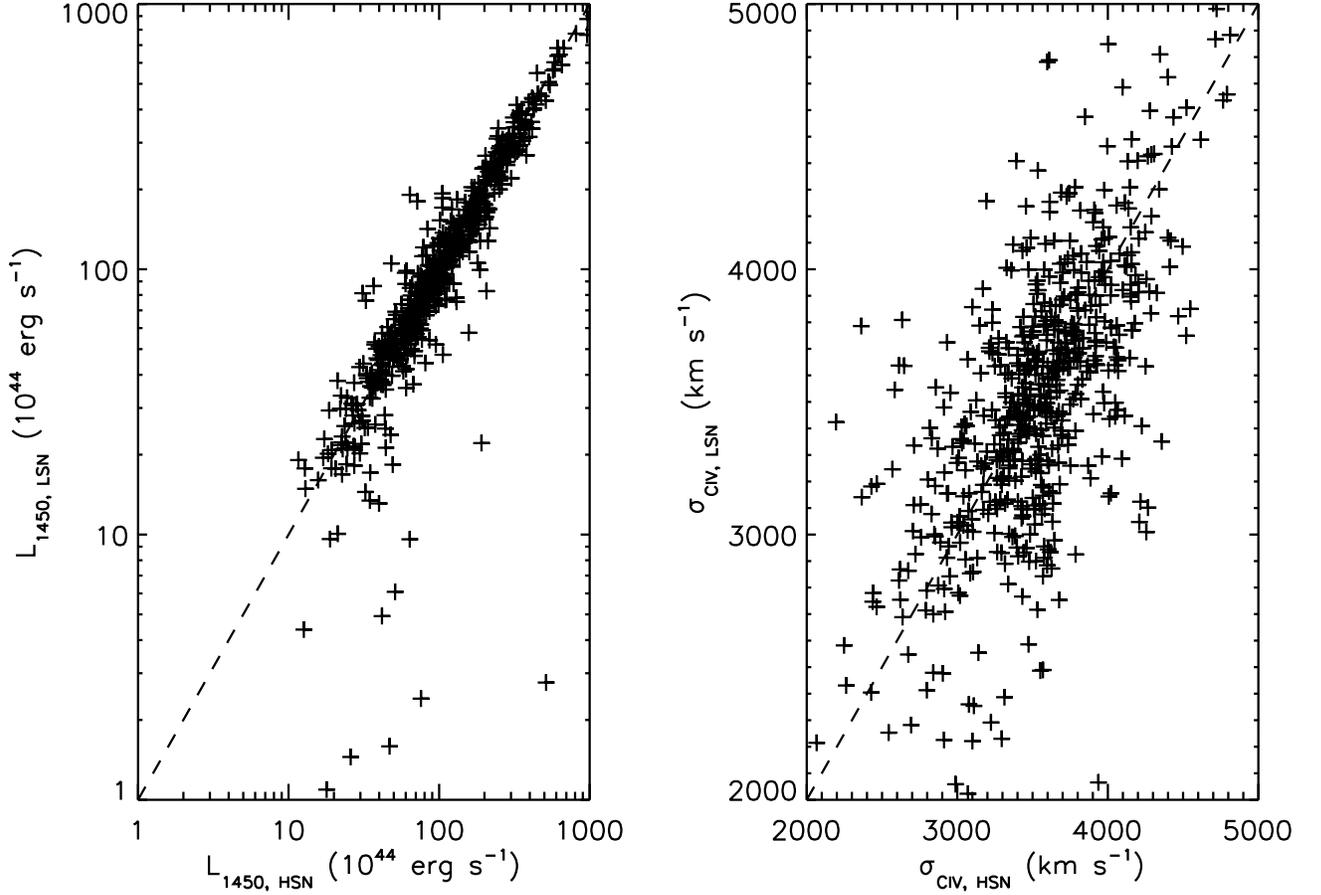}
\caption{(Left) 1450\AA\ luminosity at the high-S/N epoch (L$_{\rm 1450, HSN}$) versus the same quantity at the low-S/N epoch (L$_{\rm 1450, LSN}$). (Right) C~\textsc{iv} line dispersion at the high-S/N epoch ($\sigma_{C\textsc{iv}, {\rm HSN}}$)  versus the same quantity at the low-S/N epoch ($\sigma_{C\textsc{iv}, {\rm LSN}}$).  The dashed lines indicate zero change in the quantities between epochs.
\label{Fig4.1}}
\end{figure}

Figure \ref{Fig4.1} shows the high-S/N epoch values versus low-S/N epoch values for 1450\AA\ luminosity and C~\textsc{iv}.  The width of these distributions is due to a combination of the intrinsic variability of the quasars and the uncertainty in the measurement of those quantities.

To measure the relative change in a quantity, we will use the fractional change with respect to the average over the two epochs observed.  The fractional change in 1450\AA\ luminosity is given by Equation \ref{fracchange}:  

\begin{equation}
\Delta{\rm L}_{1450} = 2 ({\rm L}_{1450,HSN} - {\rm L}_{1450,LSN})/({\rm L}_{1450,HSN} + {\rm L}_{1450,LSN})
\label{fracchange}
\end{equation}
 $\Delta{\sigma}_{C\textsc{iv}}$ and $\Delta{\rm M}_{\rm BH}$ are defined similarly.  

\begin{figure}
\includegraphics[scale=1.0]{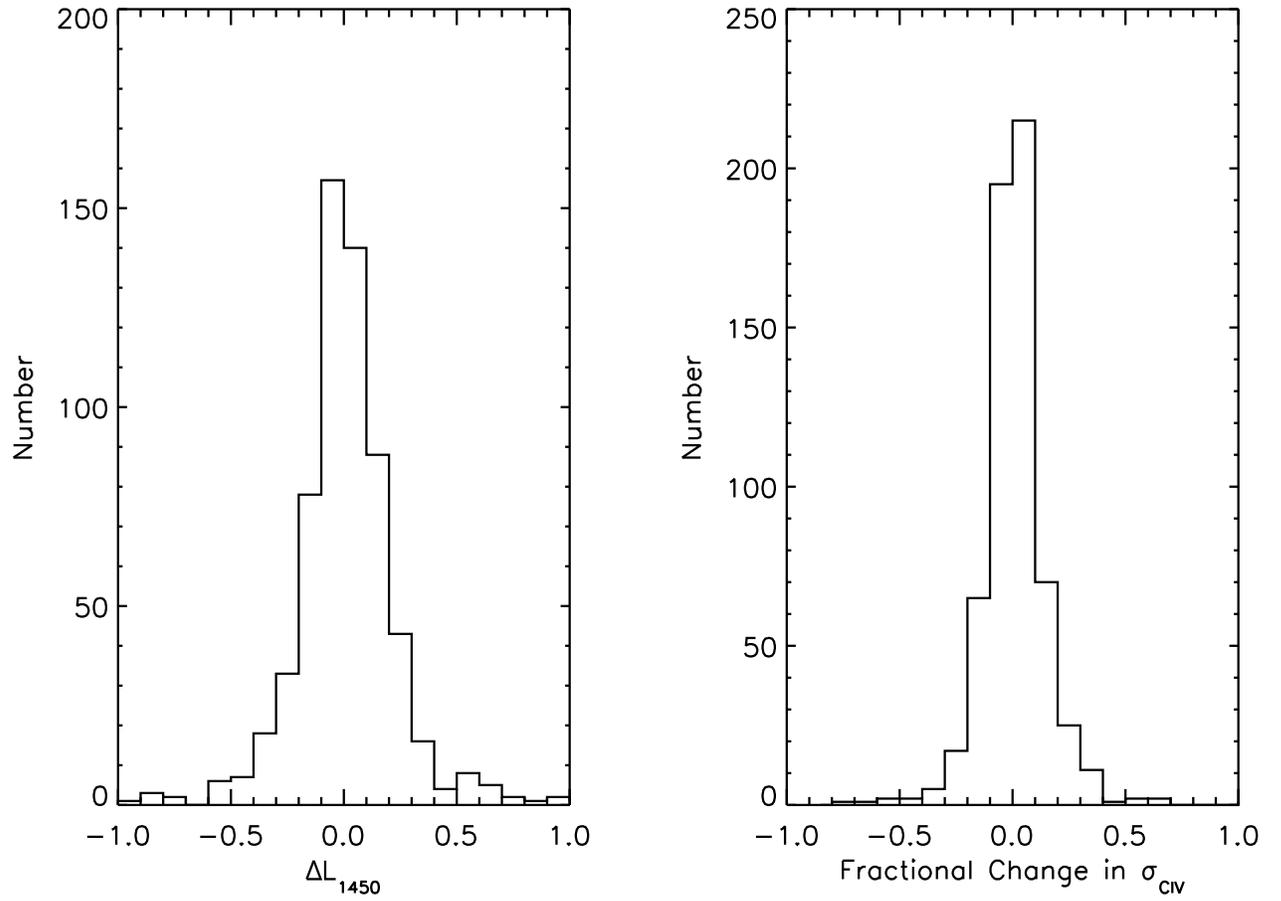}
\caption{(Left) Fractional change in 1450\AA\ luminosity.  The standard deviation of the sample is 0.161. (Right) Fractional change in C~\textsc{iv} line dispersion.  The standard deviation is 0.108.
\label{Fig4.3}} 
\end{figure}

The two panels of Figure \ref{Fig4.3} show the distribution of values of $\Delta{\rm L}_{1450}$ and $\Delta{\sigma}_{C\textsc{iv}}$.  The sample standard deviation for the $\Delta{\rm L}_{1450}$ distribution is 0.161, corresponding to a change in continuum luminosity of roughly 16\% between epochs.
%, greater than the average fractional error in L$_{1450}$ ($12\%$, as mentioned in \S\ref{luminosity}).  
%This is to be expected from a variable population---the standard deviation is a more useful statistic for measuring change than the average fractional change, which one would expect to be near zero, even for a variable population.  Thus, we will continue to use the width of multi-epoch distributions as a measure of their variability.
%Figure \ref{Fig4.2} shows the distribution of the fractional changes in ${\rm \sigma_{C\textsc{iv}}}$. 
The sample standard deviation of the ${\rm \sigma_{C\textsc{iv}}}$ distribution is 0.108, which corresponds to a change in line width of $\sim 11\%$ between epochs.
It should come as no surprise that the continuum luminosity exhibits larger variations between epochs than the line dispersion.  Much of  this variation is due to the intrinsic variability of the quasars themselves, and it is well known that quasars' continua are more variable than their emission lines (see, e.g., Paper I, Figure 13).

\begin{figure}
\includegraphics[scale=1.0]{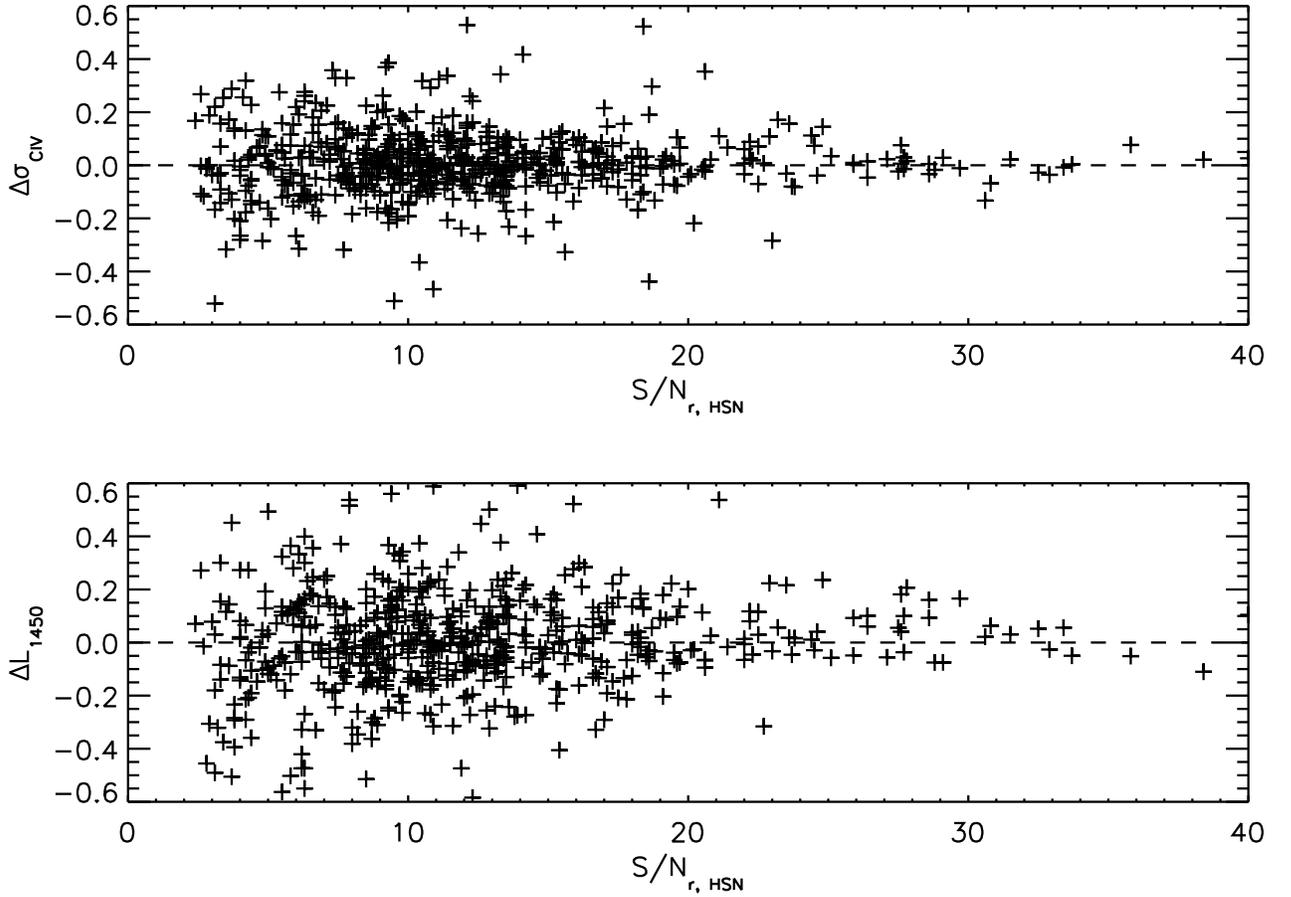}
\caption{(Upper) Fractional change in C~\textsc{iv} line dispersion as a function of $r$-band signal-to-noise ratio at the high-S/N epoch.  (Lower) Fractional change in 1450\AA\ luminosity as a function of $r$-band signal-to-noise ratio at the high-S/N epoch.  The dashed lines indicate zero change in the quantities between epochs.
\label{Fig4.5}}
\end{figure}

Figure \ref{Fig4.5} shows the fractional changes in ${\rm L}_{1450}$ and $\sigma_{C\textsc{iv}}$ as a function of high-S/N epoch signal-to-noise ratio.  The average variations are clearly, and unsurprisingly, larger for quasars with low spectral signal-to-noise ratios (${\rm S/N}_{r, {\rm HSN}} \lesssim 15$) than for quasars with high values ((${\rm S/N}_{r, {\rm HSN}} \gtrsim 15$).  However, the variations are nonzero for quasars with the highest spectral signal-to-noise ratios, an indication that intrinsic variability does play a role in these variations.

\begin{figure}
\includegraphics[scale=1.0]{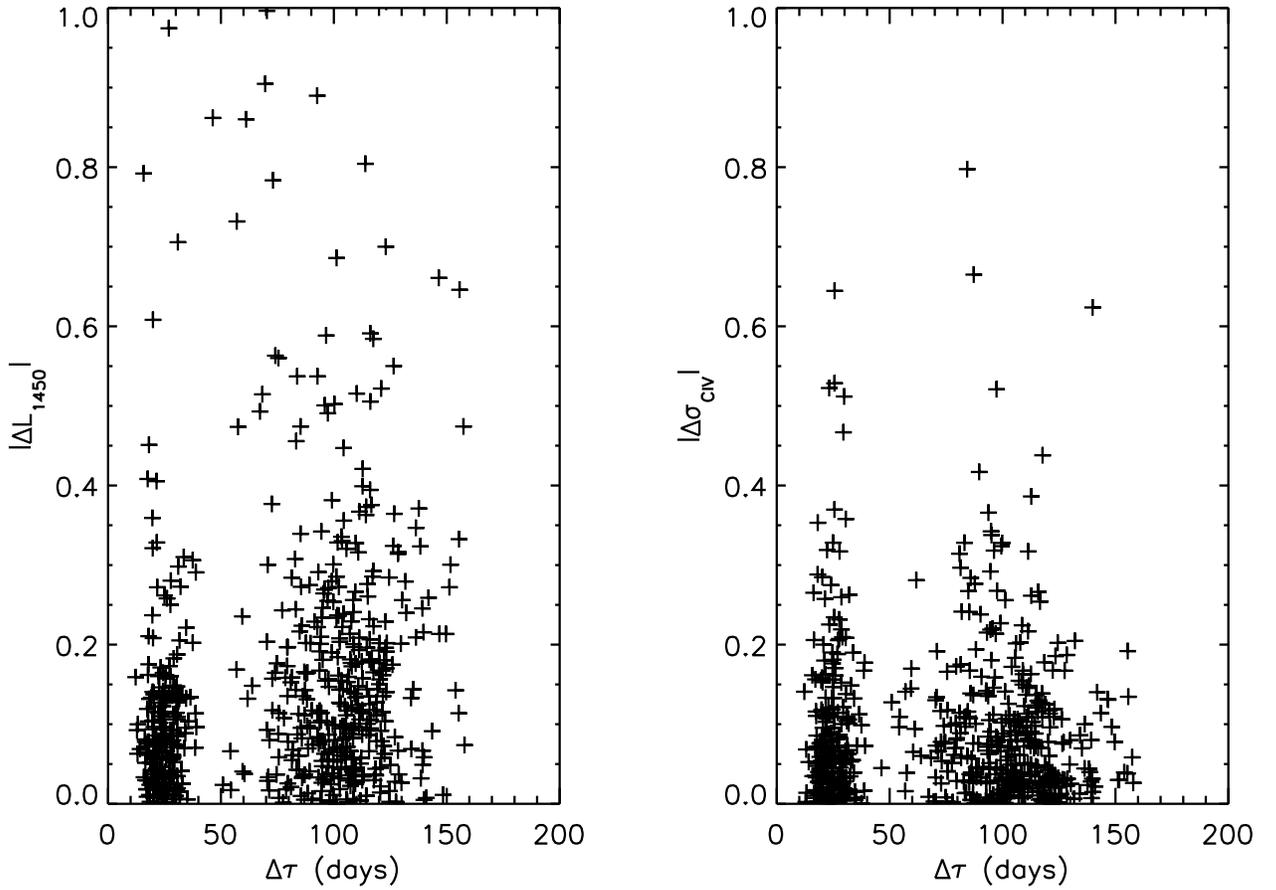}
\caption{(Left) Absolute value of the fractional change in 1450\AA\ luminosity as a function of rest-frame time lag between epochs.  (Right) Absolute value of the fractional change in C~\textsc{iv} line dispersion as a function of rest-frame time lag between epochs.  
\label{Fig4.6}}
\end{figure}

The fractional changes in continuum luminosity and C~\textsc{iv} line dispersion are shown as a function of rest-frame time lag between epochs ($\Delta{\tau}$) in Figure \ref{Fig4.6}.  To test the role of variability in these, we divide the quasars into two bins in $\Delta{\tau}$, as suggested by the distribution of observations in Figure \ref{Fig4.6}: one bin for quasars with $\Delta{\tau} <$ 50 days, and another for quasars with $\Delta{\tau} >$ 50 days.  In intrinsically time-variable populations, one would expect the variations in these quantities to show a time dependence, as seen in structure functions \citep{diclemente96,devries05}.  $\Delta{\rm L}_{1450}$ shows such a dependence.  The mean $\Delta{\rm L}_{1450}$ in the low-$\Delta{\tau}$ bin is 0.11; in the high-$\Delta{\tau}$ bin, it is 0.18.  However there is no such dependence for $\Delta{\rm \sigma_{C\textsc{iv}}}$.  The mean values of $\Delta{\sigma}_{\rm C\textsc{iv}}$ are 0.099 and 0.092 for the low- and high-$\Delta{\tau}$ bins, respectively.

This indicates that the intrinsic variability of the quasars themselves plays a larger role in the variations seen in ${\rm L}_{1450}$ than in those seen in $\sigma_{C\textsc{iv}}$.  The fact that there appears to be little difference in the size of the $\sigma_{C\textsc{iv}}$ variations between the low-$\Delta{\tau}$ and high-$\Delta{\tau}$ bins indicates that these variations are likely dominated by measurement uncertainty, not by intrinsic variability in the width of these lines.  This also suggests that the measurement errors quoted in \S\,\ref{fwhm} and Table \ref{tab2} may be underestimated.

\subsection{Variations in Estimated Black Hole Mass\label{lumfwhm_02}}

\begin{figure}
\includegraphics[scale=1.0]{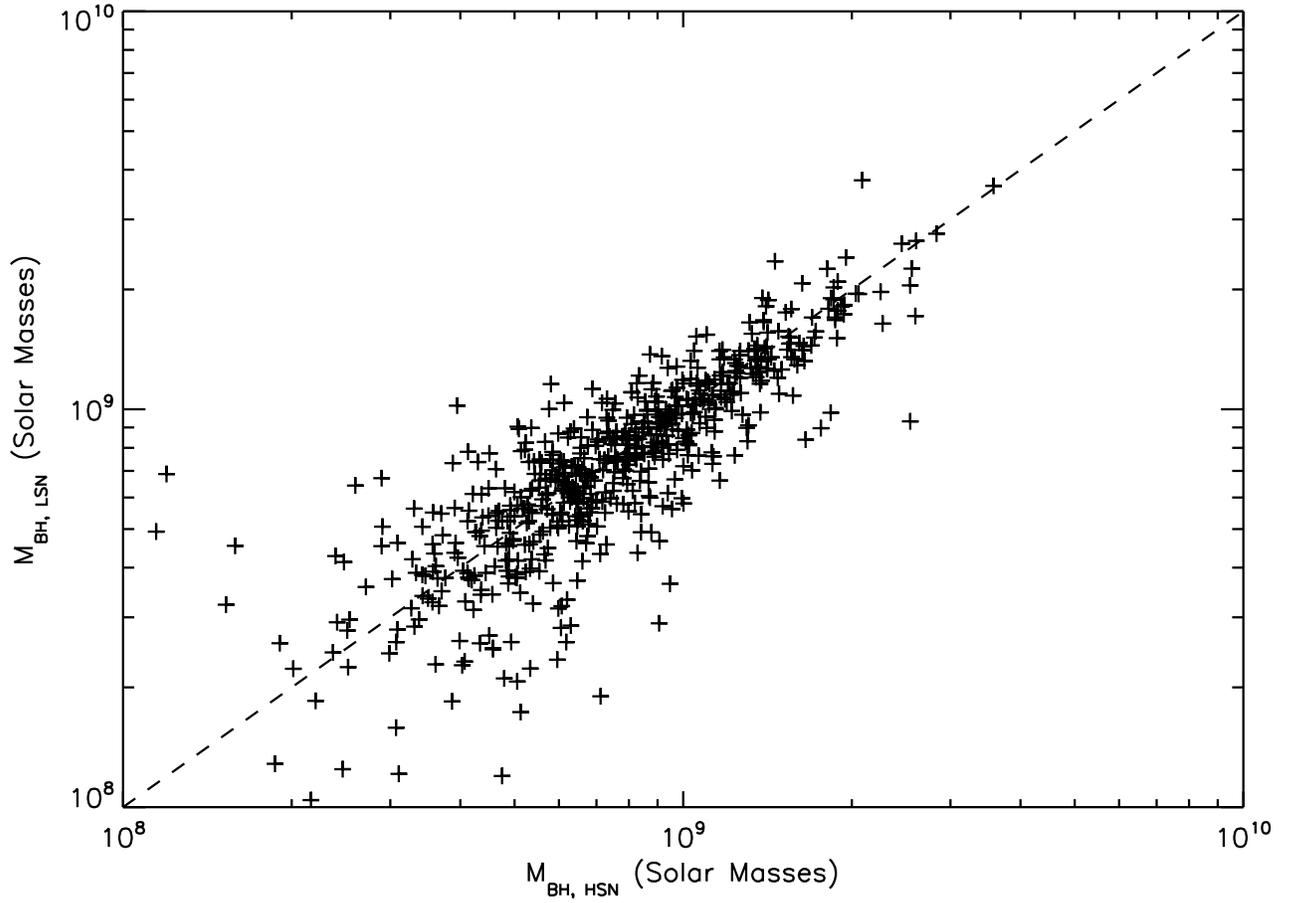}
\caption{Estimated black hole mass at the high-S/N epoch versus the same quantity at the low-S/N epoch.  The dashed line indicates zero change in M$_{\rm BH}$ between epochs.
\label{Fig4.7}}
\end{figure}

The estimate for black hole masses at the high-S/N epoch is plotted against the mass estimate from the low-S/N epoch in Figure \ref{Fig4.7}.  Most quasars do lie near the M$_{\rm BH,HSN} =$ M$_{\rm BH, LSN}$ line, indicating good general agreement in estimated mass measurements between the two epochs.

\begin{figure}
\includegraphics[scale=1.0]{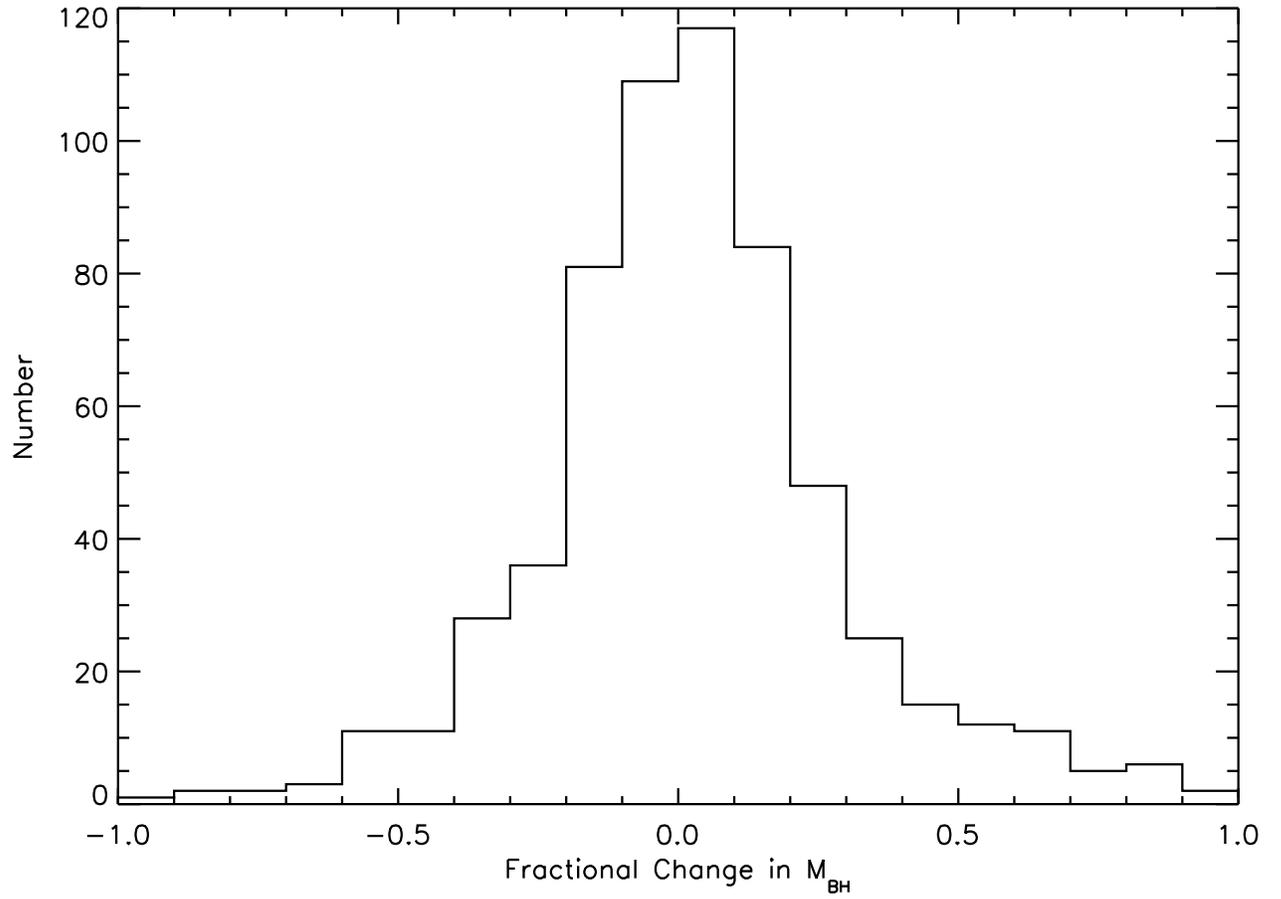}
\caption{Fractional change in estimated black hole mass.  The standard deviation of the sample is 0.301.
\label{Fig4.8}}
\end{figure}

Figure \ref{Fig4.8} shows the distribution in fractional change in the the estimate of black hole mass between epochs, $\Delta{\rm M_{BH}}$.  The standard deviation is 0.301, corresponding to a roughly $30\%$ change in the estimate between epochs.  
%The scatter of the points around a line of slope 1 (as seen in Fig. \ref{Fig4.3}) should not be interpreted as a change in the mass of the black hole between epochs.  Nor should the objects in Figure \ref{Fig4.4} demonstrating a fractional change in the black hole mass estimate between epochs be thought of as actually changing in mass.  Quasar accretion rates, on the order of a few solar masses per year  \citep[e.g.,][]{jester05}, are expected to be several orders of magnitude too low to produce a measurable change on a time scale of a year.  Instead, t
This scatter represents total inter-epoch variation in the mass estimate, due to variations in either ${\rm L}_{1450}$, ${\rm \sigma_{C\textsc{iv}}}$, or both.  Some of this change is simply due to random error in the measurements.  The rest of this scatter is due to the intrinsic variability of the quasars' luminosities and line dispersions between epochs.  
%To quantify the contribution of each of these factors, we look at the uncertainty in the fractional change in our estimate for the black hole mass.  The median uncertainty (determined by standard error propagation) in this quantity is 0.21.  Assuming that this uncertainty and inherent variability add in quadrature to yield a width of 0.30 in the $\Delta{\rm M_{BH}}$ distribution, we determine that the inherent variability in luminosity and line width alone would be responsible for a change of about 22\% in black hole mass between epochs ($0.30^{2} - 0.21^{2} = 0.22^{2}$).  Thus, the inherent variability and the measurement uncertainty contribute roughly equally to the fractional change in mass estimate between epochs.

Figure \ref{Fig4.10} shows the fractional change in estimated black hole mass as a function of the fractional change in luminosity and ${\rm \sigma_{C\textsc{iv}}}$ line dispersion.  Here it is quite clear that the line dispersion variations dominate the variations in black hole mass.  This obviously follows from Equation \ref{vestergaard}, as the mass estimate is more strongly dependent on line dispersion than continuum luminosity.  In fact, there appears to be a roughly linear relationship, with a slope of roughly 2, equal to the exponent for ${\rm \sigma_{C\textsc{iv}}}$.  However, it did not have to be the case that the line dispersion dominates the variations in black hole mass---if the variations in luminosity were much larger than those of the line dispersion, then Figure \ref{Fig4.10} might look quite different.  

In fact, given that the time delays between observations for our sample are only of the order of weeks or months in the quasars' rest frames, one would expect that the continuum luminosity variations would play a larger role in samples with longer time baselines, as structure function studies \citep{diclemente96,devries05} demonstrate that longer time baselines lead to larger average variations between observations.  

For now, we adopt $\sim 30\%$ as the contribution of inter-epoch variations to the uncertainty in the estimation of black hole masses from SDSS spectra, using the ${\rm \sigma_{C\textsc{iv}}}$ line and nearby continuum.  Given the apparent dominance of measurement uncertainty in the inter-epoch variations in the measured line dispersion, and the line dispersion's dominance of the variations in black hole mass estimate, it is not clear that we are able to set a lower limit on the effect of variability that lies below 30\%.

\begin{figure}
\includegraphics[scale=1.0]{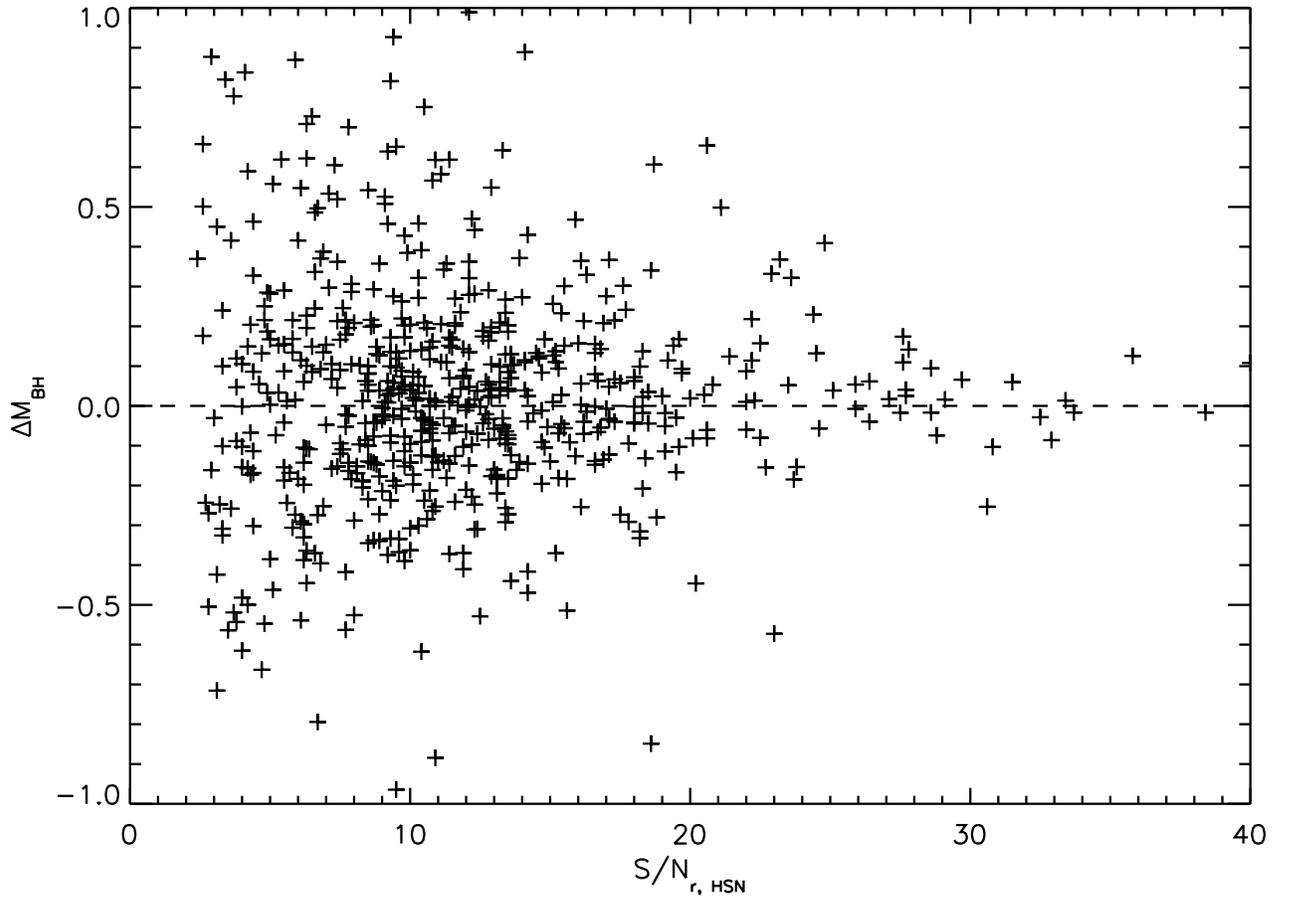}
\caption{Fractional change in estimated black hole mass as a function of $r$-band signal-to-noise ratio at the high-S/N epoch.  The dashed line indicates zero change in M$_{\rm BH}$ between epochs.
\label{Fig4.9}}
\end{figure}

\begin{figure}
\includegraphics[scale=1.0]{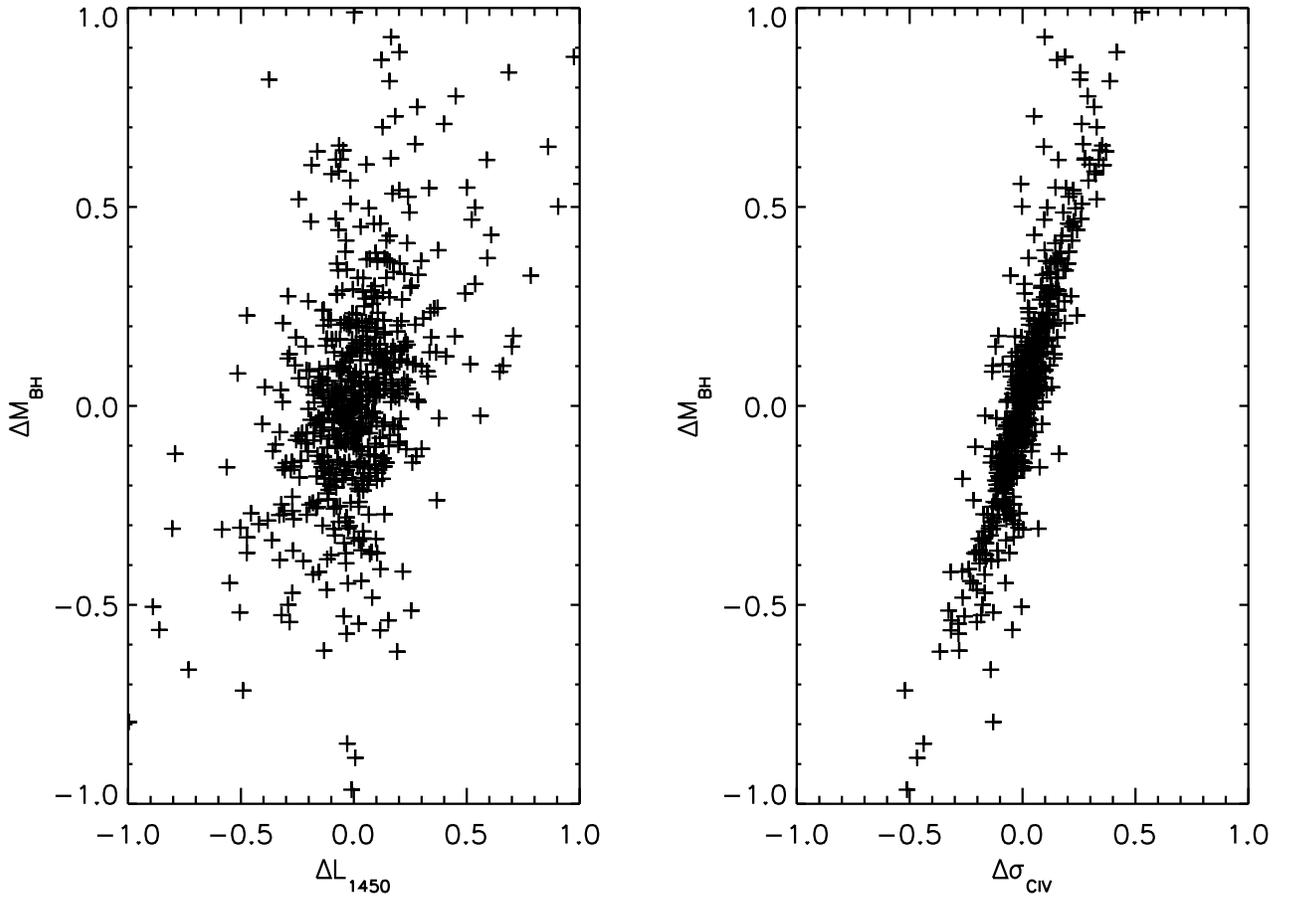}
\caption{(Left) Fractional change in estimated black hole mass as a function of the fractional change in 1450\AA\ luminosity.  (Right) Fractional change in estimated black hole mass as a function of the fractional change in the C~\textsc{iv} line dispersion.
\label{Fig4.10}}
\end{figure}

Figure \ref{Fig4.9} shows the fractional change in ${\rm M_{BH}}$ as a function of the $r$-band spectral signal-to-noise ratio.  As was the case with $\Delta{\rm L}_{1450}$ and $\Delta{\sigma}_{C\textsc{iv}}$, the width of the $\Delta{\rm M_{BH}}$ distribution decreases with increasing S/N$_{r,{\rm HSN}}$.  However, though the distribution narrows, it does not appear to be approaching zero width.

Though some of the scatter is a result of measurement uncertainties, the width of the $\Delta{\rm M_{BH}}$ distribution for high (S/N$_{r, {\rm HSN}} > 15$) signal-to-noise ratio objects does give some sense for the magnitude of the variations due to inherent quasar variability.  Thus, though it is only a rough estimate, we adopt the standard deviation of the $\Delta{\rm M_{BH}}$ distribution as a rough estimate for the size of the inter-epoch variations in estimated back hole mass.  For the 148 quasars with S/N$_{r, {\rm HSN}} > 15$, the standard deviation is 0.219, corresponding to a roughly $20\%$ change.  This is decidedly larger than the M$_{\rm BH}$ measurement uncertainty of less than $10\%$ for virtually all quasars, as seen in Figure \ref{Fig3.9} and discussed in \S\,\ref{mbhfull}

\section{Discussion\label{discussion}}

%For the most part, the change in the measured black hole masses between epochs (for the main sample and the various subsamples) is easily understood.  
Mass estimates of objects in the main sample are consistent between epochs at the $30\%$ level.  Given that the mass estimate is a function of the line width squared, but only the square root of the luminosity, it should not come as a surprise that the the variations in mass estimates are more strongly dependent on $\Delta{\rm \sigma_{C\textsc{iv}}}$ that on $\Delta{\rm L}_{1450}$.  

Even with the re-calibrated UV black hole masses, \citet{vestergaard06} find a scatter of a factor of about 2 between the UV and optical mass estimates.  Variability of the quasar continuum luminosity and line width was thought to be a possible source for this scatter.  However, this scatter is much larger than the differences in mass estimates seen between epochs in either our full sample or the quasars with signal-to-noise ratio greater than 15.
%Though this scatter comes from calibrating the relationship in Equation 2, rather than the uncertainty in the ${\rm M_{BH}}$ estimation itself, it should be encouraging that there is much less scatter between epochs than in the calibration.  
This suggests that variability is not a likely cause for the majority of this scatter.  
%In addition, it points to physical processes other than gravity playing a large role in determining the width of the C~\textsc{iv} emission line, a potentially discouraging result for the future use of C~\textsc{iv} and other high-ionization lines in estimating black hole mass.

That said, if improvements in UV techniques are possible, variability will set an ultimate limit on the precision of these techniques; it is unlikely that any estimate that relies solely on the C~\textsc{iv} emission line could do better than the $20\%$ uncertainty in ${\rm M_{BH}}$ that comes solely from the inherent variability of the continuum luminosity and C~\textsc{iv} line dispersion, as suggested by those quasars with ${\rm S/N}_{r,{\rm HSN}} > 15$.
%width.  In addition, as shown in \S\,\ref{dl1450} and \S\,\ref{dfwhm}, black hole mass estimates are even less consistent for those quasars known to be the most variable.  

%%%%%%%%%%%%%%%%%%%%%%%%%%%%%%%%
%
% Conclusions
%
%%%%%%%%%%%%%%%%%%%%%%%%%%%%%%%%

\section{Conclusions\label{conclusions}}

We have explored the effect of continuum and C~\textsc{iv} emission line variability on single-epoch estimators of quasar black hole mass.  

1) Quasar black hole mass estimates determined from SDSS spectra of the rest-frame ultraviolet show inter-epoch variations at the $30\%$ level, due to the combination of the intrinsic variability of quasars and uncertainty in the measurement of continuum luminosity and C~\textsc{iv} emission line width.

2) For our full sample, measurement error and inherent quasar variability contribute roughly equally to the inconsistencies between epochs in the estimation of ${\rm M_{BH}}$.  

3) The $\sim$20\% uncertainty in ${\rm M_{BH}}$ due to inherent variability, as suggested by the quasars with ${\rm S/N}_{r,{\rm HSN}} > 15$, sets a lower limit on the reproducibility of future UV black hole mass estimates. 
%4) Black hole mass estimates for the objects with the most variable C~\textsc{iv} emission lines are only consistent to within 53\% between epochs.  Mass estimate for quasars with the most variable continua are consistent at the $38\%$ level.  Thus, mass estimates for known highly variable quasars should be regarded as even more uncertain than those for the sample as a whole.

%5) There is no clear difference in reproducibility for those objects with the shortest and longest rest-frame time lags between observations. This is possibly due to a flat dependence in C~\textsc{iv} line variability on time, at time lags larger than light travel time from central black hole mass to broad line region.

4) Current UV black hole mass estimates for high-redshift quasars are believed to only be accurate to a factor of two, based on correlations seen with low-ionization-line mass estimates, but the smaller scatter seen here between epochs ($30\%$ for the full sample) seems to indicate that much of this scatter is yet to be understood.
%point to physical processes other gravity playing a large role in shaping the C~\textsc{iv} profile, a discouraging thought for the future use of C~\textsc{iv} in estimating black hole masses.

B.C.W. and R.J.B. would like to acknowledge support from NASA through grants NAG5-12578 and NAG5-12580, as well as support through the NSF PACI Project.  The authors made extensive use of the storage and computing facilities at the National Center for Supercomputing Applications and would like to thank the technical staff for their assistance in enabling this work.
%D.E.V. and D.P.S. wish to acknowledge the support of NSF grant AST03-07582.

Funding for the SDSS and SDSS-II has been provided by the Alfred P. Sloan Foundation, the Participating Institutions, the National Science Foundation, the U.S. Department of Energy, the National Aeronautics and Space Administration, the Japanese Monbukagakusho, the Max Planck Society, and the Higher Education Funding Council for England. The SDSS Web Site is http://www.sdss.org/.

The SDSS is managed by the Astrophysical Research Consortium for the Participating Institutions. The Participating Institutions are the American Museum of Natural History, Astrophysical Institute Potsdam, University of Basel, Cambridge University, Case Western Reserve University, University of Chicago, Drexel University, Fermilab, the Institute for Advanced Study, the Japan Participation Group, Johns Hopkins University, the Joint Institute for Nuclear Astrophysics, the Kavli Institute for Particle Astrophysics and Cosmology, the Korean Scientist Group, the Chinese Academy of Sciences (LAMOST), Los Alamos National Laboratory, the Max-Planck-Institute for Astronomy (MPIA), the Max-Planck-Institute for Astrophysics (MPA), New Mexico State University, Ohio State University, University of Pittsburgh, University of Portsmouth, Princeton University, the United States Naval Observatory, and the University of Washington.

%%%%%%%%%%%%%%%%%%%%%%%%%%%%%%%%
%
% References 
%
%%%%%%%%%%%%%%%%%%%%%%%%%%%%%%%%

\clearpage

\begin{deluxetable}{lcccccrr} 
\tablewidth{0pt}
\tablecaption{Variable Quasar Sample.  HSN and LSN indicate the high- and low-S/N ratio epochs, respectively.
\label{tab1}}
\tablehead{
\colhead{} & 
\colhead{} & 
        \multicolumn{2}{c}{MJD} & 
        \multicolumn{2}{c}{$z$} & 
	\multicolumn{2}{c}{$S/N_{r}$}\\ 
\colhead{Number} & 
\colhead{SDSS J} & 
\colhead{HSN} & 
\colhead{LSN} & 
\colhead{HSN} & 
\colhead{LSN} & 
\colhead{HSN} & 
\colhead{LSN} \\ 
}
\tablecolumns{8}
\startdata
  1 & 100543.42$-$005559.9 & 51910 & 51581 & 1.712 $\pm$ 0.002 & 1.717 $\pm$ 0.002 &  8.4 &  7.0 \\
  2 & 100356.15$-$005940.4 & 51910 & 51581 & 2.108 $\pm$ 0.002 & 2.108 $\pm$ 0.001 & 16.9 & 15.3 \\
  3 &   100013.37+011203.2 & 51910 & 51581 & 1.801 $\pm$ 0.003 & 1.805 $\pm$ 0.001 & 12.3 & 11.4 \\
  4 &   100246.85+002104.0 & 51910 & 51581 & 2.172 $\pm$ 0.002 & 2.169 $\pm$ 0.002 & 19.4 & 17.5 \\
  5 &   100412.88+001257.5 & 51910 & 51581 & 2.240 $\pm$ 0.002 & 2.243 $\pm$ 0.002 & 11.1 &  9.5 \\
  6 &   100428.43+001825.6 & 51910 & 51581 & 3.045 $\pm$ 0.001 & 2.989 $\pm$ 0.046 & 16.3 & 11.2 \\
  7 &   100623.58+004141.1 & 51910 & 51581 & 1.919 $\pm$ 0.001 & 1.916 $\pm$ 0.001 &  9.3 &  7.6 \\
  8 &   100715.53+004258.3 & 51910 & 51581 & 1.676 $\pm$ 0.001 & 1.677 $\pm$ 0.002 &  9.7 &  9.4 \\
\enddata
\end{deluxetable}

\clearpage

\begin{deluxetable}{lrrrrrr}
\tablewidth{0pt}
\tablecaption{1450\AA\ Luminosity, C~\textsc{iv} line dispersion, and estimated black hole mass at both epochs for all objects.
\label{tab2}}
\tablehead{
        \colhead{} &
        \multicolumn{2}{c}{L$_{1450}$} & 
        \multicolumn{2}{c}{$\sigma_{\rm C~\textsc{iv}}$} & 
        \multicolumn{2}{c}{log(M$_{BH}$/M$_{\sun}$)} \\
        \colhead{} &
        \multicolumn{2}{c}{$10^{44}$ erg s$^{-1}$} & 
        \multicolumn{2}{c}{km s$^{-1}$} & 
        \colhead{} &
        \colhead{} \\
        \colhead{Number} &
        \colhead{HSN} &
        \colhead{LSN} & 
        \colhead{HSN} & 
        \colhead{LSN} & 
        \colhead{HSN} & 
        \colhead{LSN}
}
\tablecolumns{7}
\startdata
  1 &   48 $\pm$   8 &   57 $\pm$  11 & 3834 $\pm$   97 &  3740 $\pm$  159 & 8.79 $\pm$ 0.05 & 8.81 $\pm$ 0.06 \\
  2 &  197 $\pm$  17 &  212 $\pm$  22 & 4197 $\pm$  125 &  4408 $\pm$  122 & 9.19 $\pm$ 0.03 & 9.25 $\pm$ 0.03 \\
  3 &  105 $\pm$  10 &  193 $\pm$  18 & 3456 $\pm$  141 &  3440 $\pm$  261 & 8.88 $\pm$ 0.04 & 9.01 $\pm$ 0.07 \\
  4 &  396 $\pm$  23 &  315 $\pm$  22 & 3371 $\pm$  118 &  3320 $\pm$  119 & 9.16 $\pm$ 0.03 & 9.10 $\pm$ 0.04 \\
  5 &  208 $\pm$  21 &  164 $\pm$  20 & 3942 $\pm$  144 &  3970 $\pm$  159 & 9.15 $\pm$ 0.04 & 9.10 $\pm$ 0.05 \\
  6 &  379 $\pm$  45 &  268 $\pm$  49 & 2919 $\pm$  114 &  2708 $\pm$  174 & 9.03 $\pm$ 0.04 & 8.88 $\pm$ 0.07 \\
  7 &   86 $\pm$  11 &   73 $\pm$  10 & 3294 $\pm$  181 &  2229 $\pm$  475 & 8.79 $\pm$ 0.06 & 8.41 $\pm$ 0.19 \\
  8 &   58 $\pm$   8 &   48 $\pm$   6 & 3784 $\pm$  189 &  3904 $\pm$  155 & 8.82 $\pm$ 0.06 & 8.81 $\pm$ 0.05 \\
\enddata
\end{deluxetable}

%\onecolumn
\clearpage

%%%%%%%%%%%%%%%%%%%%%%%%%%%%%%%%
%
%Figures
%
%%%%%%%%%%%%%%%%%%%%%%%%%%%%%%%%

% use scale = 0.5, 0.3, 0.3 respectively

%Section 2 Figures

\clearpage

\clearpage

\clearpage

%Section 3 Figures

\clearpage

%;\begin{figure}
%\includegraphics[scale=1.0]{f4.eps}
%\includegraphics[scale=1.0]{alll1450unchist.eps}
%\caption{Uncertainty in the 1450\AA\ luminosity at the high-S/N (dark histogram) and low-S/N (gray histogram) epochs.
%\label{Fig3.2}}
%\end{figure}
%\clearpage

%\begin{figure}
%\includegraphics[scale=1.0]{f5.eps}
%\includegraphics[scale=1.0]{rsnl1450unc.eps}
%\caption{(Left) Uncertainty in the 1450\AA\ luminosity as a function of $r$-band signal-to-noise ratio (${\rm S/N}_{r}$) at the high-S/N epoch.  (Right) The same quantities at the low-S/N epoch.
%\label{Fig3.3}}
%\end{figure}
%\clearpage

\clearpage

%\begin{figure}
%\includegraphics[scale=1.0]{f7.eps}
%\includegraphics[scale=1.0]{allcivsigunchist.eps}
%\caption{Uncertainty in the C~\textsc{iv} line dispersion at the high-S/N (dark histogram) and low-S/N (gray histogram) epochs.
%\label{Fig3.5}}
%\end{figure}
%\clearpage

%\begin{figure}
%\includegraphics[scale=1.0]{f8.eps}
%\includegraphics[scale=1.0]{rsnfwhmunc.eps}
%\caption{(Left) Uncertainty in the C~\textsc{iv} line dispersion as a function of $r$-band signal-to-noise ratio at the high-S/N epoch.  (Right) The same quantities at the low-S/N epoch.
%\label{Fig3.6}}
%\end{figure}
%\clearpage

\clearpage

%\begin{figure}
%\includegraphics[scale=1.0]{f10.eps}
%\includegraphics[scale=1.0]{allmbh1unchist.eps}
%\caption{Logarithm of the uncertainty in the estimated black hole mass at the high-S/N (dark histogram) and low-S/N (gray histogram) epochs.
%\label{Fig3.8}}
%\end{figure}
%\clearpage

\clearpage

%Section 4 Figures

\clearpage

%\begin{figure}
%\includegraphics[scale=1.0]{f13.eps}
%\includegraphics[scale=1.0]{allcivsig12.eps}
%\caption{C~\textsc{iv} line dispersion at the high-S/N epoch ($\sigma_{C\textsc{iv}, HSN}$)  versus the same quantity at the low-S/N epoch ($\sigma_{C\textsc{iv}, LSN}$).
%\label{Fig4.2}}
%\end{figure}
%\clearpage

\clearpage

%\begin{figure}
%\includegraphics[scale=1.0]{f15.eps}
%\includegraphics[scale=1.0]{alldcivsighist.eps}
%\caption{Fractional change in C~\textsc{iv} line dispersion.  The standard deviation of the sample is 0.108.
%\label{Fig4.4}}
%\end{figure}
%\clearpage

\clearpage

\clearpage

\clearpage

\clearpage

\clearpage

%\begin{figure}
%\includegraphics[scale=1.0]{f21.eps}
%\includegraphics[scale=1.0]{alldtaudmbh.eps}
%\caption{Fractional change in estimated black hole mass as a function of rest-frame time lag between epochs.
%\label{Fig4.10}}
%\end{figure}
%\clearpage

\clearpage

%%%%%%%%%%%%%%%%%%%%%%%%%%%%%%%%
%
%Tables
%
%%%%%%%%%%%%%%%%%%%%%%%%%%%%%%%%

%%\clearpage
%%Table 1. -- Plate List

%%\clearpage

\end{document}